\documentclass[prb,num, preprint, showpacs, 10pt, twocolumn,superscriptaddress]{revtex4}
\usepackage[utf8]{inputenc}
\usepackage[T1]{fontenc}
\usepackage[colorlinks=true, pdfstartview=FitV, linkcolor=blue, citecolor=blue]{hyperref}
\usepackage{color}
\usepackage{epstopdf}
\usepackage{graphicx}
\usepackage{amsmath}
\usepackage{array,multirow}
\usepackage{tabularx}
\makeatletter
\def\hlinewd#1{%
\noalign{\ifnum0=`}\fi\hrule \@height #1 %
\futurelet\reserved@a\@xhline}

\begin{document}


\preprint{}

\title{Theoretical study of phase transitions in Sb$_2$S$_3$, Bi$_2$S$_3$ and Sb$_2$Se$_3$ under compression}

\author{E. Lora da Silva}
\email{esdasil@idf.upv.es}
\affiliation{Instituto de Dise\~{n}o para la Fabricaci\'{o}n y Producci\'{o}n Automatizada, MALTA Consolider Team, Universitat Polit\`{e}cnica de Val\`{e}ncia, Val\`{e}ncia, Spain}
%\affiliation{I3N and Department of Physics, University of Aveiro, 3810-093 Aveiro, Portugal}

%\author{M. C. Santos}
%\affiliation{CFisUC, Department of Physics, University of Coimbra, P-3004-516 Coimbra, Portugal}
%\date{\today}

\author{J. M. Skelton}
\affiliation{School of Chemistry, University of Manchester, Oxford Road, Manchester M13 9PL, United Kingdom}

\author{P. Rodr\'{i}guez-Hern\'{a}ndez}
\author{A. Mu\~{n}oz}
\affiliation{Departamento de F\'{i}sica, Instituto de Materiales y Nanotecnología, MALTA Consolider Team, Universidad de La Laguna, Tenerife, Spain}

\author{D. Mart\'{i}nez-Garc\'{i}a}
\affiliation{Departamento de F\'{i}sica Aplicada - ICMUV, MALTA Consolider Team, Universitat de Val\`{e}ncia}

\author{F. J. Manj\'{o}n}
\affiliation{Instituto de Dise\~{n}o para la Fabricaci\'{o}n y Producci\'{o}n Automatizada, MALTA Consolider Team, Universitat Polit\`{e}cnica de Val\`{e}ncia, Val\`{e}ncia, Spain}

%\email[]{a.walsh@imperial.ac.uk}
\date{\today}

\begin{abstract}
We report a theoretical study of Sb$_2$S$_3$, Sb$_2$Se$_3$ and Bi$_2$S$_3$ sesquichalcogenides at hydrostatic pressures up to 60 GPa. We explore the possibility that the \textit{R-3m}, \textit{C2/m}, \textit{C2/c} and the disordered \textit{Im-3m} phases observed in sesquichalcogenides with heavier cations, viz. Bi$_2$Se$_3$, Bi$_2$Te$_3$ and Sb$_2$Te$_3$, could also be formed in Sb$_2$S$_3$, Sb$_2$Se$_3$ and Bi$_2$S$_3$, as suggested by recent experiments. Our calculations show that the \textit{C2/m} and \textit{C2/c} phases are energetically unstable for any of the three compounds over the entire range of pressures examined. In contrast, the disordered bcc-like \textit{Im-3m} phase is energetically stable at high pressures; however, it is only for Sb$_2$Se$_3$ that the disordered phase presents dynamical stability below 60 GPa. Our calculations further show that at ambient pressure the \textit{Pnma} phase is the most energetically favourable for Sb$_2$S$_3$ and Bi$_2$S$_3$ whereas, and surprisingly, for Sb$_2$Se$_3$ it is the \textit{R-3m} phase which presents the lowest enthalpy energy at 0 GPa, in contradiction to experimental evidence. From lattice dynamics and elastic tensor calculations we observe that both \textit{Pnma} and \textit{R-3m} phases are dynamically and mechanically stable at 0 GPa. These results suggest that the formation of the \textit{R-3m} phase for Sb$_2$Se$_3$ could be feasible at close to ambient conditions. Furthermore, and to aid the identification of this phase, we provide the theoretical crystal structure (lattice and atomic parameters) and complete infrared and Raman spectra.
\end{abstract}

%\pacs{}% insert suggested PACS numbers in braces on next line

\maketitle %\maketitle must follow title, authors, abstract and \pacs

\date{\today}

%\pacs{61.72.Bb, 61.80.Az, 71.55.Cn, 71.70.Ej}

%%%%%%%%%%%%%%%%%%%%%%%%%%%%%%%%%%%%
\section{Introduction}
%%%%%%%%%%%%%%%%%%%%%%%%%%%%%%%%%%%%

Since the identification of the trigonal tetradymite-like \textit{R-3m} phases of group-15 sesquichalcogenides (i.e. Sb$_2$Te$_3$, Bi$_2$Se$_3$, Bi$_2$Te$_3$) as 3D topological insulators,\cite{Science.325.178.2009, NatPhys.5.438.2009} the family of A$_2$X$_3$ sesquichalcogenides has attracted a great deal of attention from the scientific community. Three-dimensional topological insulators represent a new class of matter, with insulating bulk electronic states and topologically-protected metallic surface states due to time-reversal symmetry and strong spin-orbit coupling, and present potential interest for spintronics and quantum computing applications.\cite{RevModPhys.282.3045.2010} Due to this fundamental interest and potential applications, identifying new topological insulators and materials with superconducting properties is currently among the most widely-studied topics in condensed matter science.

\begin{figure*}
\begin{center}
\includegraphics[width=12cm]{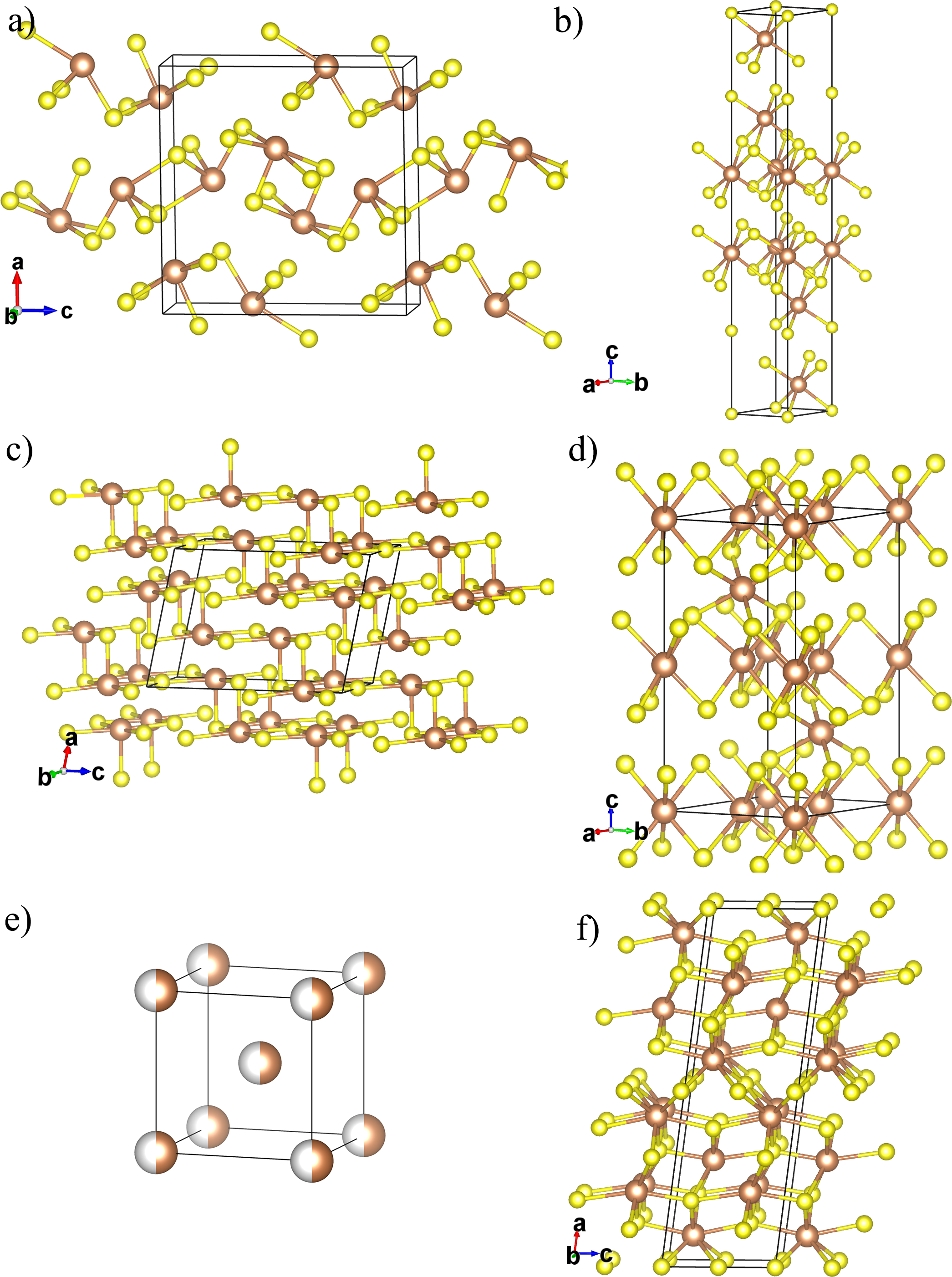}
\caption{\label{fig:unit-cell}
Images of the \textit{Pnma} (a), \textit{R-3m} (b), \textit{C2/m} (c), \textit{C2/c} (d), disordered bcc-type \textit{Im-3m} (e) A$_2$X$_3$ sesquichalcogenide structures (A = Sb, Bi; X = S, Se). The \textit{C2/m} nine/ten-fold structure used to model the disordered \textit{Im-3m} phase is also shown for comparison (f). The A cations and X anions are shown as brown and yellow spheres, respectively.} 
\end{center}
\end{figure*}

Stibnite (Sb$_2$S$_3$), bismuthinite (Bi$_2$S$_3$) and antimonselite (Sb$_2$Se$_3$) minerals are also group-15 sesquichalcogenides; however, they do not crystallize at room conditions in the tetradymite-like structure, but in the orthorhombic U$_2$S$_3$-type (\textit{Pnma}) structure (Fig. \ref{fig:unit-cell}.a). Sb$_2$S$_3$, Bi$_2$S$_3$ and Sb$_2$Se$_3$ are semiconductors with band-gap widths of 1.7, 1.3, and 1.2 eV, respectively.\cite{JPhysChemSolids.2.240.1957,PhysRevB.87.205125.2013} These materials are used in a wide range of technological applications including photovoltaic solar cells, X-ray computed tomography detectors, fuel cells, gas sensors and for detection of biomolecules.\cite{JPhysChemLett.1.1524.2010,AdvFunctMater.21.4663.2011,NatPhotonics.9.409.2015,NatEnergy.2.17046.2017,NatMater.5.118.2006,PhysChemB.110.21408.2006,NanoLett.9.1482.2009}

The \textit{Pnma} structure has been identified as a possible post-post-perovskite phase of (Mg,Fe)SiO$_3$ minerals and of NaFeN$_3$ at high pressure (HP).\cite{arxiv.1410.2783.2014, MineralMRCag.80.659.2016} Thus, the study of Sb$_2$S$_3$, Bi$_2$S$_3$ and Sb$_2$Se$_3$ at HP could also provide useful information about the HP behaviour of the ABO$_3$ minerals, which are found in the mantle of the Earth. In this context, initial experimental HP studies of Sb$_2$S$_3$, Bi$_2$S$_3$ and Sb$_2$Se$_3$ have shown that the \textit{Pnma} structure is stable under compression, with first-order phase transitions (PTs) occurring around 50 GPa.\cite{SciRep.3.2665.2013,JPhysChemA.118.1713.2014,JPhysCondMatter.28.015602.2016,JPhysChemC.120.10547.2016,JAlloysCompd.688.329.2016} Curiously enough, recent HP studies have found that Sb$_2$Se$_3$ becomes a topological superconductor at around 10 GPa and 2.5 K,\cite{SciRep.4.6679.2014} exhibiting highly conducting spin-polarized surface states similar to Bi$_2$Se$_3$. \cite{PhysRevB.97.235306.2018} Moreover, three further studies have suggested that several first- and second-order PTs occur for Sb$_2$S$_3$ up to 50 GPa.\cite{SciRep.6.24246.2016,PhysRevB.97.024103.2018,SciRep.8.14795.2018} Furthermore, it has also been suggested that the HP phases of Sb$_2$S$_3$ could be similar to those observed for heavier sesquichalcogenides such as Bi$_2$Se$_3$, Bi$_2$Te$_3$ and Sb$_2$Te$_3$.\cite{PhysStatusSolidiB.250.669.2013} Therefore, it is of interest to examine and compare the stability of different structural phases (\textit{R-3m}, \textit{C2/m}, \textit{C2/c} and disordered \textit{Im-3m}) observed for heavier cation sesquichalcogenides (i.e. Bi$_2$Se$_3$, Bi$_2$Te$_3$ and Sb$_2$Te$_3$) on our three minerals of interest, Sb$_2$S$_3$, Bi$_2$S$_3$ and Sb$_2$Se$_3$, and at different pressure conditions.

On the other hand, several theoretical studies performed on the \textit{R-3m} structure of Sb$_2$Se$_3$ have suggested that this phase should undergo a topological quantum phase transition under compression. \cite{PhysRevB.84.245105.2011,PhysRevB.89.035101.2014} In fact, it has been reported that such a topological transition was observed experimentally at ~2 GPa.\cite{PhysRevLett.110.107401.2013} Furthermore, recent calculations suggest that the tetradymite-like Sb$_2$Se$_3$ structure becomes a topological insulator at ambient pressure.\cite{PhysRevB.97.075147.2018} Consequently, HP studies performed on these group-15 sesquichalcogenides are highly relevant to the research on topological states, therefore possible HP phases of these compounds should be thoroughly evaluated. It is worthy of mentioning, that a recent work performed on Bi$_2$S$_3$ predicts the system to be unstable under compression, and decomposing into another stoichiometric system \cite{JPhysChemLett.9.5785.2018}

In this work, we report theoretical simulations at 0 K of the \textit{Pnma} and hypothetical \textit{R-3m}, \textit{C2/m}, \textit{C2/c} and \textit{Im-3m} phases for Sb$_2$S$_3$, Sb$_2$Se$_3$ and Bi$_2$S$_3$ (Figure \ref{fig:unit-cell}), with a view to assessing which, if any, are likely to fulfil either of the stability conditions under hydrostatic pressure.

%%%%%%%%%%%%%%%%%%%%%%%%%%%%%%%%%%%%
\section{Theoretical Methodology}
%%%%%%%%%%%%%%%%%%%%%%%%%%%%%%%%%%%%

The structural properties of the different crystalline phases of Sb$_2$S$_3$, Bi$_2$S$_3$ and Sb$_2$Se$_3$ were calculated within the framework of density-functional theory (DFT).\cite{hohenberg-pr-136-1964} The Vienna \textit{Ab-initio} Simulation Package (VASP) code \cite{kresse-cms-6-1996} was employed to perform simulations with the projector augmented-wave (PAW) scheme including six valence electrons for S[3s$^2$3p$^4$] and Se[4s$^2$4p$^4$] and fifteen valence electrons for Sb[4d$^10$5s$^2$5p$^3$] and Bi[5d$^10$6s$^2$6p$^3$]. Convergence of the total energy was achieved with a plane-wave kinetic-energy cut-off of 600 eV. The generalized-gradient approximation (GGA) functional with the Perdew-Burke-Ernzerhof parameterization revised for solids (PBEsol), \cite{perdew-prl-100-2008,perdew-prl-102-2009} was used for all the calculations. 

The Brillouin-zone (BZ) was sampled with $\Gamma$-centered Monkhorst-Pack \cite{monkhorst-prb-13-1976} grids employing adequate meshes for the different structural phases of the three compounds: \textit{Pnma} - 6 $\times$ 10 $\times$ 6, \textit{R-3m} - 12 $\times$ 12 $\times$ 12, \textit{C2/m} - 6 $\times$ 12 $\times$ 6, \textit{C2/c} - 10 $\times$ 10 $\times$ 8, and \textit{Im-3m} (using a \textit{C2/m} conventional cell) - 6 $\times$ 12 $\times$ 12.  

The disordered \textit{Im-3m} phase is a body-centered cubic (bcc) disordered structure, and has been theoretically predicted and experimentally found for Bi$_2$Te$_3$ in 2011. \cite{PhysRevLett.106.145501.2011}
For sesquichacogenides with A$_2$X$_3$ stoichiometry, the bcc lattice site (2a Wyckoff position) is randomly occupied by 40\% of A cations and 60\% of X anions. This means that such a structure is a disordered phase with a mixture of cations and anions randomly sharing the same bcc crystallographic position and forming a A-X substitutional alloy.\cite{PhysRevLett.106.145501.2011} Due to the theoretical difficulty in simulating the disordered \textit{Im-3m} structure, we have used a 9/10-fold \textit{C2/m} structure (formation of 9/10 chemical A-X bonds), as was previously employed for Bi$_2$Te$_3$ \cite{PhysRevLett.106.145501.2011} and Bi$_2$Se$_3$.\cite{JPhysChemC.117.10045.2013} Moreover it has been observed that the 9/10-fold \textit{C2/m} structure presents a bcc-like structural order, in agreement with the observed XRD patterns; \cite{PhysRevLett.106.145501.2011, SciRep.8.14795.2018} therefore giving support to employ the calculated intermediate bcc-like monoclinic \textit{C2/m} phase to confirm the experimental presence of the disordered \textit{Im-3m} phase.

Structural relaxations were performed by allowing the atomic positions and the unit-cell parameters to change during the ionic relaxation, at different volume values (compressions). From these we obtain the respective external pressure for the specific applied compression (isotropic volume compression) and hence respective set of crystal structures. The pressure-volume (P-V) curves for all the compounds were fitted to a third-order Birch-Murnaghan equation of state \cite{murnaghan-pnas-30-1944,birch-pr-71-1947} to obtain the equilibrium volume, bulk modulus and, respective pressure derivative. The enthalpy, $H$, curves were computed by considering the relation, $H=E+pV$, where $E$ is the total electronic energy of the system, $p$ is pressure, and $V$ is the volume. The analysis and comparison of the $H$ curves for the different polymorphs can provide insights regarding the thermodynamic stability of each phase for increasing pressure values, up until the studied pressure range (60 GPa).

Lattice-dynamics calculations were performed for the energetically favourable polymorphs at different pressure values, namely the \textit{Pnma} and \textit{R-3m} phases of Sb$_2$Se$_3$ at room pressure, and for the disordered \textit{Im-3m} phases of all three compounds, as explained in detail in Secs. \ref{subsec:thermo} and \ref{subsec:phonon}. The phonon properties were computed by using the supercell finite-displacement method implemented in the Phonopy package \cite{togo-prb-78-2008} with VASP used as the force calculator. \cite{chaput-prb-84-2001} Supercells were expanded up to 2 $\times$ 4 $\times$ 2 for the \textit{Pnma} phase, and 2 $\times$ 2 $\times$ 2 for the \textit{R-3m} and disordered phases; to allow the exact calculation of frequencies at the zone center ($\Gamma$) and inequivalent zone-boundary wavevectors, which were then interpolated to obtain phonon-dispersion curves and density of states on a uniform 50 $\times$ 50 $\times$ 50 $\Gamma$-centered \textbf{q}-point mesh.

To correct for the long-range Coulomb interaction (LO-TO splitting), a non-analytical correction, based on the Born effective-charge tensors and the electronic-polarization component of the macroscopic static dielectric tensor, was applied.\cite{PhysRevB.50.13035R.1994,PhysRevB.55.10355.1997} These quantities were obtained using the density-functional perturbation theory (DFPT) method implemented in VASP.\cite{gajdos-prb-73-2006}

Infrared (IR) and Raman spectra were calculated for the ground-state \textit{R-3m} phase of the Sb$_2$Se$_3$ structure by employing the methods described in Ref. \onlinecite{PhysChemChemPhys.19.12452.2017} and implemented in the Phonopy-Spectroscopy package.\cite{PhonoptSpec} The linewidths were obtained by computing the third-order force constants and following the many-body perturbative approach described in detail in Refs. \onlinecite{PhysChemChemPhys.19.12452.2017} and \onlinecite{PhysRevB.91.094306.2015} and implemented in the Phono3py software.\cite{PhysRevB.91.094306.2015}

Elastic tensors were computed to assess the mechanical stability of the two energetically favourable phases of Sb$_2$Se$_3$ at 0 GPa, namely the \textit{Pnma} and the \textit{R-3m} polymorphs. Respective calculations were carried out by employing the central-difference method, where the unique components of the elastic tensor are determined by performing six finite distortions of the lattice and deriving the tensor elements from the strain-stress relationship.\cite{PhysRevB.65.104104.2002} For these calculations, it was necessary to increase the plane-wave energy cutoff to converge the stress tensor adequatly, which was achieved by systematically increasing the plane-wave cutoff up until 950 eV. We then further employed the ELATE software \cite{JPhysCondensMatter.28.275201.2016} to analyse the linear compressibility using the computed stress tensors.

%%%%%%%%%%%%%%%%%%%%%%%%%%%%%%%%%%%%
\section{Results and Discussion}
%%%%%%%%%%%%%%%%%%%%%%%%%%%%%%%%%%%%
%%%%%%%%%%%%%%%%%%%%%%%%%%%%%%%%%%%%
\subsection{Structural properties of the \textit{Pnma} phase}
%%%%%%%%%%%%%%%%%%%%%%%%%%%%%%%%%%%%

The \textit{Pnma} phase of the A$_2$X$_3$ structures are composed by weak stacking interactions which hold the layers along the $a$-axis together, which description becomes challenging for conventional DFT functionals.\cite{JSolidStateChem.213.116.2014,ChemSci.6.5255.2015,JPhysChemC.120.10547.2016} We have compared the equilibrium lattice parameters, bulk moduli and pressure derivatives calculated for the \textit{Pnma} phases of Sb$_2$Se$_3$, Sb$_2$S$_3$ and Bi$_2$S$_3$, with existing experimental and theoretical results found in literature (Tab. \ref{table:pnma_param}), to verify the accuracy of our theoretical calculations as a prior step before attempting the study of the HP phases. 

\begingroup
\squeezetable
\begin{table}
%\begin{center}
\caption{ \label{table:pnma_param} 
Calculated equilibrium lattice parameters (a$_0$, b$_0$ and c$_0$), equilibrium bulk moduli (B$_0$) and pressure derivatives (B$_0'$), at 0 GPa, of the \textit{Pnma} phase of Sb$_2$Se$_3$, Sb$_2$S$_3$ and Bi$_2$S$_3$. Values are compared to experimental and other theoretical results found in literature.}
%\begin{ruledtabular}
\begin{tabular}{|c |c|c|c|c|c|c |}\hlinewd{1pt}
 & \multicolumn{2}{c|}{\textbf{Sb$_2$Se$_3$}}  & \multicolumn{2}{c|}{\textbf{Sb$_2$S$_3$}} & \multicolumn{2}{c|}{\textbf{Bi$_2$S$_3$ }}   \\ \hline
 \multirow{5}{*}{\textbf{a$_0$ (\AA)}} & \multicolumn{2}{c|}{11.75$^a$}  & \multicolumn{2}{c|}{11.24$^a$} & \multicolumn{2}{c|}{11.19$^a$}   \\ \cline{2-7}
 &  \textit{Theo.} &	\textit{Exp.} & \textit{Theo.} &	\textit{Exp.} & \textit{Theo.} &	\textit{Exp.}\\\cline{2-7}
 & 11.80$^b$ &  11.80$^{f}$  &  11.27$^b$  	&   11.30$^{b,j,k}$  &  11.41$^b$  &  11.27$^p$ \\
 & 11.52$^c$  &   11.79$^g$   &  11.02$^c$  	&    11.31$^{l,m}$   &  11.00$^{n}$  & 11.33$^q$ \\
 & 11.91$^d$ &				&	11.30$^h$	&				     &	11.58$^{o}$	&			\\
 & 11.53$^e$ &				 &	11.08$^i$	&				     &				&			\\\hline
 \multirow{5}{*}{\textbf{b$_0$ (\AA)}} & \multicolumn{2}{c|}{3.98$^a$}  & \multicolumn{2}{c|}{3.83$^a$} & \multicolumn{2}{c|}{3.96$^a$}   \\ \cline{2-7}
 &   \textit{Theo.} &	\textit{Exp.} & \textit{Theo.} &	\textit{Exp.} & \textit{Theo.} &	\textit{Exp.}\\\cline{2-7}
  & 3.99$^{b}$  &  3.98$^{f}$  &  3.81$^c$  & 3.84$^{b,j,k}$  &  3.97$^b$  &  3.97$^p$ \\
  & 3.96$^{c,e}$ & 3.99$^{g}$ &  3.84$^h$  & 	3.84$^{l,m}$ 	&  3.94$^{n}$  & 3.98$^q$ 		 \\
  & 3.98$^d$   &			   &	3.83$^{b,i}$	   &				    &	3.99$^{o}$	&    	\\\hline
  \multirow{5}{*}{\textbf{c$_0$ (\AA)}} & \multicolumn{2}{c|}{11.30$^a$}  & \multicolumn{2}{c|}{10.91$^a$} & \multicolumn{2}{c|}{10.94$^a$}   \\ \cline{2-7}
 &   \textit{Theo.} &	\textit{Exp.} & \textit{Theo.} &	\textit{Exp.} & \textit{Theo.} &	\textit{Exp.}\\\cline{2-7}
  & 11.28$^b$   &  11.65$^{f,g}$  &  10.89$^b$  & 11.23$^{b,j,l,m}$  &  11.01$^b$  &  11.13$^p$ \\
  & 11.22$^{c,e}$  &  	 	    &  10.79$^c$  &  11.24$^{k}$ 		& 	10.83$^{n}$	  & 11.18$^q$  \\
  & 11.70$^d$ 	&				&	11.22$^h$&				       &	11.05$^{o}$	&\\
  &				&				&	10.81$^i$&				      &				& \\ \hline\hline
     \multirow{5}{*}{\textbf{V$_0$ (\AA$^3$)}} & \multicolumn{2}{c|}{528.11$^a$}  & \multicolumn{2}{c|}{469.6$^a$} & \multicolumn{2}{c|}{484.4$^a$}   \\ \cline{2-7}
 &   \textit{Theo.} &	\textit{Exp.} & \textit{Theo.} &	\textit{Exp.} & \textit{Theo.} &	\textit{Exp.}\\\cline{2-7}
  & 531.1$^b$  &  547.1$^{f}$    &  470.4$^b$  & 486.0$^{b}$      &  498.3$^b$  &  498.4$^p$ \\
  &  511.8$^c$  &  547.5$^g$    &   453.0$^c$  & 487.7$^{i,m,j}$    & 469.1$^{n}$ &  501.6$^q$ \\
  &  598.1$^r$  & 552.5$^s$    	&   529.9$^r$  & 488.2$^{k}$      & 510.1$^{o}$    &            \\
    &  			 & 			  	&   			 & 				    & 511.6$^q$    &            \\
\hline
    \multirow{5}{*}{\textbf{B$_0$ (GPa)}} & \multicolumn{2}{c|}{31.1$^a$}  & \multicolumn{2}{c|}{31.5$^a$} & \multicolumn{2}{c|}{42.3$^a$}   \\ \cline{2-7}
 &   \textit{Theo.} &	\textit{Exp.} & \textit{Theo.} &	\textit{Exp.} & \textit{Theo.} &	\textit{Exp.}\\\cline{2-7}
  &  70.5$^c$ &  30.0$^{f}$  &  32.2$^b$  & 37.6$^{b}$   &  	83.6$^{n}$   &  36.6$^p$	 \\
  &          &   32.7$^{s}$ &    80.3$^c$  &  26.9$^{j}$ & 	32.3$^{o}$  &  38.9$^q$	 \\
  &  		&				&			& 27.2$^k$		&	 36.5$^q$   	& 37.5$^u$ 	\\
    &  		&				&			& 41.4$^t$		&		     	&			\\\hline
     \multirow{5}{*}{\textbf{B$_0'$ }} & \multicolumn{2}{c|}{6.6$^a$}     & \multicolumn{2}{c|}{6.6$^a$} & \multicolumn{2}{c|}{6.8$^a$}   \\ \cline{2-7}
 &   \textit{Theo.} &	\textit{Exp.} & \textit{Theo.} &	\textit{Exp.} & \textit{Theo.} &	\textit{Exp.}\\\cline{2-7}
  &           &  6.1$^{f}$  &  6.2$^b$  & 3.8$^{b}$    & 5.9$^q$  		&  6.4$^p$  \\
  &          &  5.6$^{s}$ &  		  & 7.9$^{j}$      &  6.4$^{o}$ 	&	5.5$^q$	\\
    &         & 			&  	       & 6.0$^{k}$     &  			 &	4.6$^u$	\\
    &  		&				&		& 7.8$^t$	  &	       		   &		\\\hline
\hlinewd{1pt}
\end{tabular}
\footnotemark[1]{This work,} %a
\footnotemark[2]{Ref. \onlinecite{JPhysChemC.120.10547.2016},} %b
\footnotemark[3]{Ref. \onlinecite{SolidStateSci.14.1211.2012},}  %c
\footnotemark[4]{Ref. \onlinecite{EJChem.6.S147.2009},} %d
\footnotemark[5]{Ref. \onlinecite{ChemSci.6.5255.2015},} %e
\footnotemark[6]{Ref. \onlinecite{SciRep.3.2665.2013},}  %f
\footnotemark[7]{Ref. \onlinecite{ZKristall.171.261.1985},}  %g
\footnotemark[8]{Ref. \onlinecite{PhysicaB.406.287.2011},}  %h
\footnotemark[9]{Ref. \onlinecite{PhysChemChemPhys.16.345.2014},}  %i
\footnotemark[10]{Ref. \onlinecite{PhysChemMinerals.30.463.2003},}  %j
\footnotemark[11]{Ref. \onlinecite{SciRep.6.24246.2016},}  %k
\footnotemark[12]{Ref. \onlinecite{ZKristall.135.308.1972},}  %l
\footnotemark[13]{Ref. \onlinecite{PhysChemMiner.135.29.254.2002}, } %m
\footnotemark[14]{Ref. \onlinecite{JMolModel.20.2180.2014},}  %n
\footnotemark[15]{Ref. \onlinecite{CompMatSci.101.301.2015},} %o
\footnotemark[16]{Ref. \onlinecite{PhysChemMinerals.32.578.2005},}  %p
\footnotemark[17]{Ref. \onlinecite{JPhysChemA.118.1713.2014},}  %q
\footnotemark[18]{Ref. \onlinecite{JSolidStateChem.213.116.2014},}  %r
\footnotemark[19]{Ref. \onlinecite{SciRep.4.6679.2014},}  %s
\footnotemark[20]{Ref. \onlinecite{SciRep.8.14795.2018},}  %t
\footnotemark[21]{Ref. \onlinecite{JAlloysCompd.688.329.2016}}  %u
%\end{ruledtabular}
%\end{center}
\end{table}
\endgroup

By observing Tab. \ref{table:pnma_param} we may find that our calculated lattice paramters of Sb$_2$Se$_3$, calculated at room pressure (a$_0$ = 11.75~\AA, b$_0$ = 3.98~\AA~ and c$_0$ = 11.30~\AA), are found to be in good agreement with experimental values detailed in Refs. \onlinecite{SciRep.3.2665.2013} and \onlinecite{ZKristall.171.261.1985} (a$_0$ = 11.80~\AA, b$_0$ = 3.97~\AA, c$_0$ = 11.65~\AA~ and  a$_0$=11.79~\AA, b$_0$ = 33.98~\AA~ and c$_0$ = 11.65~\AA, respectively), and also with values obtained from \textit{ab initio} calculations found in literature. \cite{JPhysChemC.120.10547.2016, SolidStateSci.14.1211.2012, EJChem.6.S147.2009,ChemSci.6.5255.2015} We must mention that the most notable deviation of our calculated values from experimental measurements is a $\sim$3\% reduction of the c$_0$ parameter (not surprisingly leading to an underestimation of the V$_0$ of roughly $\sim$3-4\% when compared to experiment). In fact our results are comparable to the referenced theoretical results of Ref. \onlinecite{JPhysChemC.120.10547.2016}, where calculations were also carried out by employing PAW-PBEsol. We must also note that the c$_0$ obtained from Ref. \onlinecite{EJChem.6.S147.2009} of 11.70~\AA~is quite high when compared to the present results, however closer to experimental results; whereas the a$_0$ parameter has a larger error when compared to experimental values. This fact has to do with the use of the GGA-PBE functional, which is well known to overestimate volumes (therefore overestimating the volumes up to $\sim$10\%, as can be confirmed in Ref. \onlinecite{JSolidStateChem.213.116.2014}). On the other hand, LDA underestimates volumes, as observed from the lattice parameters provided in Refs. \onlinecite{ChemSci.6.5255.2015}, \onlinecite{SolidStateSci.14.1211.2012} and \onlinecite{JMolModel.20.2180.2014}. Interestingly enough we must note that along the $b$-axis, the difference between theoretical and experimental values are very small, since this is the crystallographic direction where covalently bonded chains prevail.

Similar results were obtained for our calculated lattice parameters of Sb$_2$S$_3$. As shown on Tab. \ref{table:pnma_param}, the calculated parameters for this compound agrees with experimental measurements \cite{JPhysChemC.120.10547.2016, PhysChemMinerals.30.463.2003} as well as with other theoretical results.\cite{JPhysChemC.120.10547.2016,SolidStateSci.14.1211.2012,PhysicaB.406.287.2011,PhysChemChemPhys.16.345.2014} We must note however, and as already described for Sb$_2$Se$_3$, that the lattice parameters obtained from LDA calculations tend to underestimate respective values, and therefore the discrepancy found for a$_0$ presented from Refs. \onlinecite{SolidStateSci.14.1211.2012} and \onlinecite{PhysChemChemPhys.16.345.2014}. The $c_0$ parameter is however closer to our PBEsol results than the GGA-PBE values described in Ref. \onlinecite{PhysicaB.406.287.2011}, which actually provides a better agreement to experimental results.\cite{JPhysChemC.120.10547.2016,PhysChemMinerals.30.463.2003} %although affecting the B$_0$ as already mentioned for Sb$_2$Se$_3$. \onlinecite{PhysChemChemPhys.16.345.2014}

Results obtained for the Bi$_2$S$_3$ system is also consistent with several data found in literature, both experimental as well as theoretical. We note that calculations performed on Bi$_2$S$_3$ by employing the Armiento and Mattsson 2005 parametrized GGA functional (AM05)\cite{JChemPhys.128.084714.2008,PhysRevB.72.085108.2005,PhysRevB.79.155101.2009} seems to show a slightly better reproduction of c$_0$ with respect to experimental measurements.\cite{JPhysChemC.120.10547.2016}

The last two rows of Tab. \ref{table:pnma_param} show the calculated B$_0$ and B$_0’$ results of the \textit{Pnma} phase of the three compounds and respective comparison to experimental measurements and other calculations.
The calculated values obtained by fitting the P-V curves of Sb$_2$Se$_3$ to a third-order Birch-Murnaghan equation are B$_0$ = 31.1 GPa (B$_0'$ = 6.6), which is close to the experimental values of B$_0$ = 30 GPa (B$_0'$ = 6.1) from Ref. \onlinecite{SciRep.3.2665.2013} and  B$_0$ = 32.7 GPa (B$_0'$ = 5.6) from Ref. \onlinecite{CompMatSci.101.301.2015}.

%SciRep.3.2665.2013,ZKristall.171.261.1985,SciRep.4.6679.2014} 

For Sb$_2$S$_3$, we have obtained B$_0$ = 31.5 GPa (B$_0'$ = 6.6), which is within the range of experimental values, \cite{JPhysChemC.120.10547.2016,PhysChemMinerals.30.463.2003,SciRep.6.24246.2016,SciRep.8.14795.2018} and also with PAW-PBEsol calculations of Ref. \onlinecite{JPhysChemC.120.10547.2016}. Moreover, these results are also close to those experimentally measured for the As-doped stibnite mineral.\cite{HighTempHighPress.43.351.2014} 

Finally, our values for Bi$_2$S$_3$ result in B$_0$ = 42.3 GPa (B$_0'$ = 6.8), which is consistent with DFT data \cite{CompMatSci.101.301.2015,JPhysChemA.118.1713.2014} and experimental values reported in Refs. \onlinecite{PhysChemMinerals.32.578.2005}, \onlinecite{JPhysChemA.118.1713.2014} and \onlinecite{JAlloysCompd.688.329.2016}.

\subsection{Energetic Stability}
\label{subsec:thermo}

\begin{figure}
\begin{center}
\includegraphics[width=8cm]{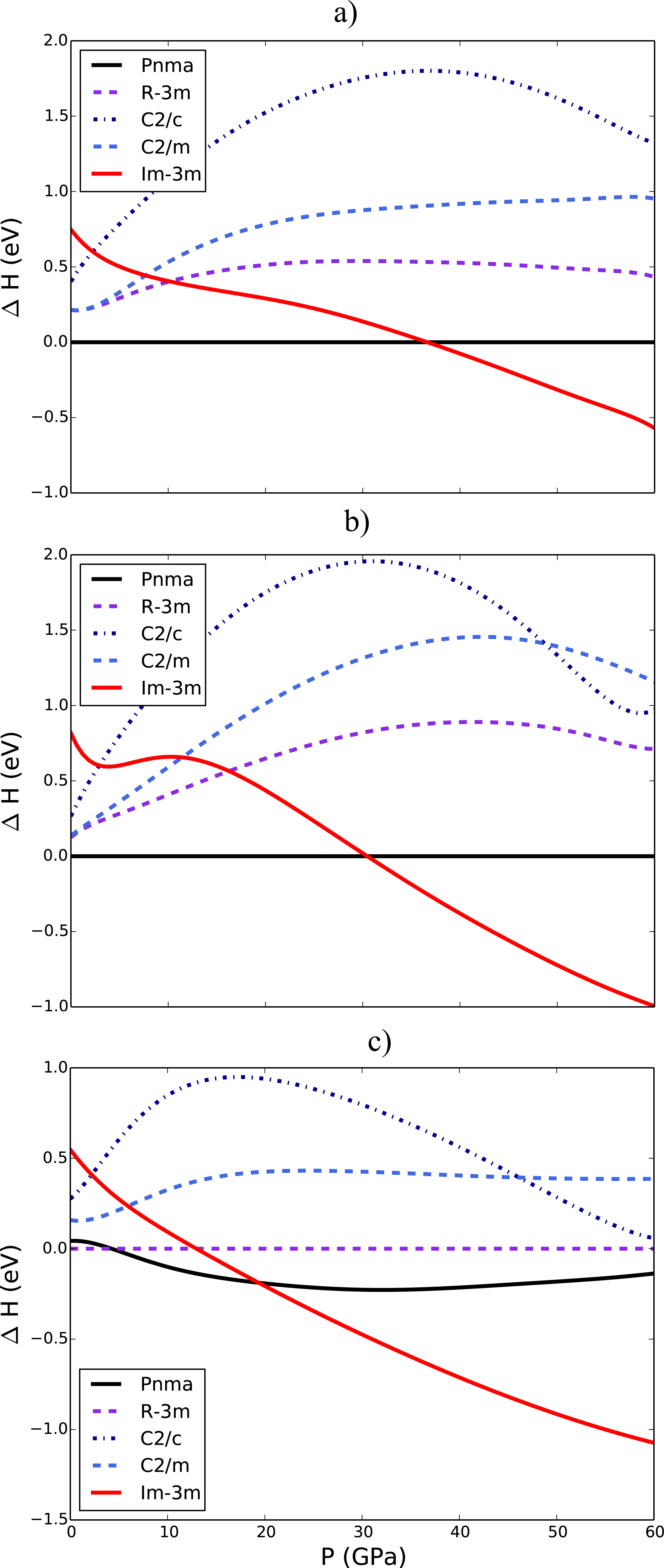}
\caption{\label{fig:enthalpy}
Calculated enthalpy \textit{vs} pressure curves, for the different possible phases (shown in Fig \ref{fig:unit-cell}) of Sb$_2$S$_3$ (a), Bi$_2$S$_3$ (b) and Sb$_2$Se$_3$ (c), relative to the lowest-energy phase at ambient pressure: the \textit{Pnma} phase for Sb$_2$S$_3$ and Bi$_2$S$_3$ and the \textit{R-3m} phase for Sb$_2$Se$_3$.} 
\end{center}
\end{figure}

Since our calculations on the \textit{Pnma} phases were found to be in good agreement with the overall data found in literature, we have proceeded in carrying out a theoretical study of the hypothetical \textit{R-3m}, \textit{C2/m}, \textit{C2/c} and disordered \textit{Im-3m} phases of Sb$_2$S$_3$, Bi$_2$S$_3$ and Sb$_2$Se$_3$ to probe whether such phases could be energetically competitive under hydrostatic pressure. Figs. \ref{fig:enthalpy}a, \ref{fig:enthalpy}b and \ref{fig:enthalpy}c show the pressure-dependence of the enthalpy differences, relative to the stable phase at ambient pressure, between the five above mentioned phases of Sb$_2$S$_3$, Bi$_2$S$_3$ and Sb$_2$Se$_3$, respectively. Values of the predicted transition pressures between the different phases are summarized in Tab. \ref{table:transition}.

\begin{table}
\begin{center}
\caption{ \label{table:transition} 
Theoretical estimation of the pressure-induced phase transitions of \textit{R-3m} $\rightarrow$ \textit{Pnma} and \textit{Pnma} $\rightarrow$ disordered \textit{Im-3m} for the Sb$_2$Se$_3$, Sb$_2$S$_3$ and Bi$_2$S$_3$ compounds (presented in units of GPa).}
\begin{tabular}{| c |c|c|c|}\hlinewd{1pt}
  & \textbf{Sb$_2$Se$_3$}  & \textbf{Sb$_2$S$_3$} & \textbf{Bi$_2$S$_3$ }  \\ \hline
\textbf{\textit{R-3m} }$\rightarrow$ \textbf{\textit{Pnma}} & 4.78  & -- & --   \\ \hline
 \textbf{\textit{Pnma} }$\rightarrow$ \textbf{\textit{Im-3m}} & 21.07  & 35.12 & 30.08   \\ 
\hlinewd{1pt}
\end{tabular}
\end{center}
\end{table}

From the enthalpy plots (Fig. \ref{fig:enthalpy}) we may observe that :
\begin{enumerate}
\item At 0 GPa the orthorhombic \textit{Pnma} phase is energetically stable for Bi$_2$S$_3$ and Sb$_2$S$_3$; however for Sb$_2$Se$_3$ it is the trigonal \textit{R-3m} phase the most favourable phase at 0 GPa.
\item The two monoclinic \textit{C2/c} and \textit{C2/m} phases do not become energetically competitive with the ground-state phase, over the range of pressures examined, in any of the three compounds. 
\item The bcc-like disordered \textit{Im-3m} structure, which can be understood as a disordered solid solution of atoms, is the most energetically stable phase at pressures above 35, 30 and 21 GPa for Sb$_2$S$_3$, Bi$_2$S$_3$ and Sb$_2$Se$_3$, respectively.
\end{enumerate}

With respect to the first point, referring to Bi$_2$S$_3$ and Sb$_2$S$_3$, our calculations predict that the \textit{Pnma} structure is energetically the most stable phase throughout the whole range of studied pressures, as expected from experimental evidences that show the observation of respective phase, both at ambient and at high pressure. Surprisingly, however, our simulations indicate that the \textit{R-3m} phase of Sb$_2$Se$_3$ is the most stable phase at pressures below 4.8 GPa, being both \textit{Pnma} and \textit{R-3m} phases energetically competitive between 0 and 4.8 GPa. This feature contradicts the experimental findings of Sb$_2$Se$_3$ consistently crystallizing to the \textit{Pnma} phase at ambient conditions. We must note however that at 0 K and 0 GPa, the energy difference between the two phases is only 22.71 meV (per f.u.), which is lower than the thermal barrier (k$_\mathrm{B}$T $\sim$ 25 meV at 300 K) required for the phase transition to occur under ambient conditions. In order to probe whether the vibrational contributions to the free energy could alter the energy ordering between the two phases, we have further plotted the free energies, where the entropy terms are obtained from our lattice dynamics calculations, and which will be discussed in more detail in Sec. \ref{subsec:phonon}.

Regarding the second point, our analysis further shows that for the three studied compunds, the two monoclinic \textit{C2/c} and \textit{C2/m} phases are never energetically competitive up until 60 GPa. These results are compatible with the fact that no phase transition had previously been observed in Sb$_2$S$_3$, Bi$_2$S$_3$ or Sb$_2$Se$_3$ under compression up to $\sim$50 GPa.\cite{SciRep.4.6679.2014,SciRep.3.2665.2013,JPhysChemA.118.1713.2014,JPhysCondMatter.28.015602.2016,JPhysChemC.120.10547.2016,JAlloysCompd.688.329.2016} However, three recent studies have reported low-pressure phase transitions occuring for Sb$_2$S$_3$.\cite{SciRep.6.24246.2016,PhysRevB.97.024103.2018,SciRep.8.14795.2018} A transition to an unknown phase was observed around 15 GPa,\cite{SciRep.6.24246.2016,PhysRevB.97.024103.2018} and several other transitions were also evidenced between 10 and 25 GPa, and tentatively proposed to be the \textit{R-3m}, \textit{C2/c} and \textit{C2/m} structural phases. It must however be clarified that this latter study applied an ethanol-methanol mixture as the pressure-transmitting medium, therefore there is a possibility that the observed transitions could have been induced by non-hydrostatic pressure effects.
 
Finally, as for the third point, from a thermodynamic point-of-view our results indicate that the bcc-like disordered \textit{Im-3m} phase, initially identified for Bi$_2$Se$_3$, Bi$_2$Te$_3$ and Sb$_2$Te$_3$,\cite{PhysRevLett.106.145501.2011,JPhysChemC.117.10045.2013,InorChem.50.11291.2011} seem to be energetically favourable at HP for our three materials of interest. These results are consistent with the observation of such a phase at around 50 GPa for Sb$_2$Se$_3$\cite{SciRep.3.2665.2013} and above 25 GPa for Sb$_2$S$_3$.\cite{SciRep.6.24246.2016,PhysRevB.97.024103.2018,SciRep.8.14795.2018} However, for the Bi$_2$S$_3$ structure, our results do not agree with those found in Refs.~\onlinecite{JPhysChemA.118.1713.2014} and \onlinecite{JAlloysCompd.688.329.2016}, where disorder has been observed above 50 GPa although attributed mostly to a pressure-induced amorphization. Moreover, a more recent work claims that Bi$_2$S$_3$ is unstable above 31.5 GPa, decomposing into a mixture of BiS$_2$ and BiS compounds.\cite{JPhysChemLett.9.5785.2018}

In summary, the agreement of our results regarding the observation of the disordered \textit{Im-3m} phase for Sb$_2$Se$_3$ and Sb$_2$S$_3$, but not for Bi$_2$S$_3$, suggests that thermodynamic stability is not sufficient to explain the lack of the HP disordered phase for the latter compound. In the following section we discuss the dynamical stability of the \textit{Im-3m} phase as a function of pressure in order to provide a deeper understanding regarding this question.

\subsection{Dynamical Stability}
\label{subsec:phonon}

Energetic stability is a necessary, but not sufficient condition for a structural phase to be synthetically accessible. One should also probe the dynamical stability of the system, which requires the study of the phonon frequencies. If imaginary frequencies emerge (usually represented by negative frequencies in the phonon dispersion curves), this would indicate that the system is at a potential-energy maximum (transient state), undergoing a phase transition and thus cannot be kinetically stable at the given temperature and/or pressure conditions.\cite{ProcCambridgePhilosSoc36.160.1940,Dove.IntLattDyn,Dove.StrutDyn,AmMin.82.213.1997, BullMaterSci.1.129.1979, PhysRevLett.111.025503.2013}

In this section, we consider the phonon properties of different phases for the three compounds, which were observed to be energetically the most favourable (Fig. \ref{fig:enthalpy}) at different pressure values, namely: 
\begin{enumerate}
\item The disordered \textit{Im-3m} phases of the three materiales at different pressure ranges.
\item The \textit{Pnma} phase of the three systems at 50 GPa.
\item The \textit{Pnma} and \textit{R-3m} phases of Sb$_2$Se$_3$ at 0 GPa.
\end{enumerate}

\subsubsection{The Disordered BCC-Type \textit{Im-3m} Phase}
To assess the possibility of dynamical stability for the disordered \textit{Im-3m} phases of the three compounds at HP, we have evaluated the phonon dispersion curves at pressure values of 30 GPa, which is close to the transition pressures observed in Fig. \ref{fig:enthalpy}; and at higher pressures of 50 (Sb$_2$Se$_3$) and 60 GPa (Sb$_2$S$_3$, Bi$_2$S$_3$). 

\begin{figure*}
\begin{center}
\includegraphics[width=12cm]{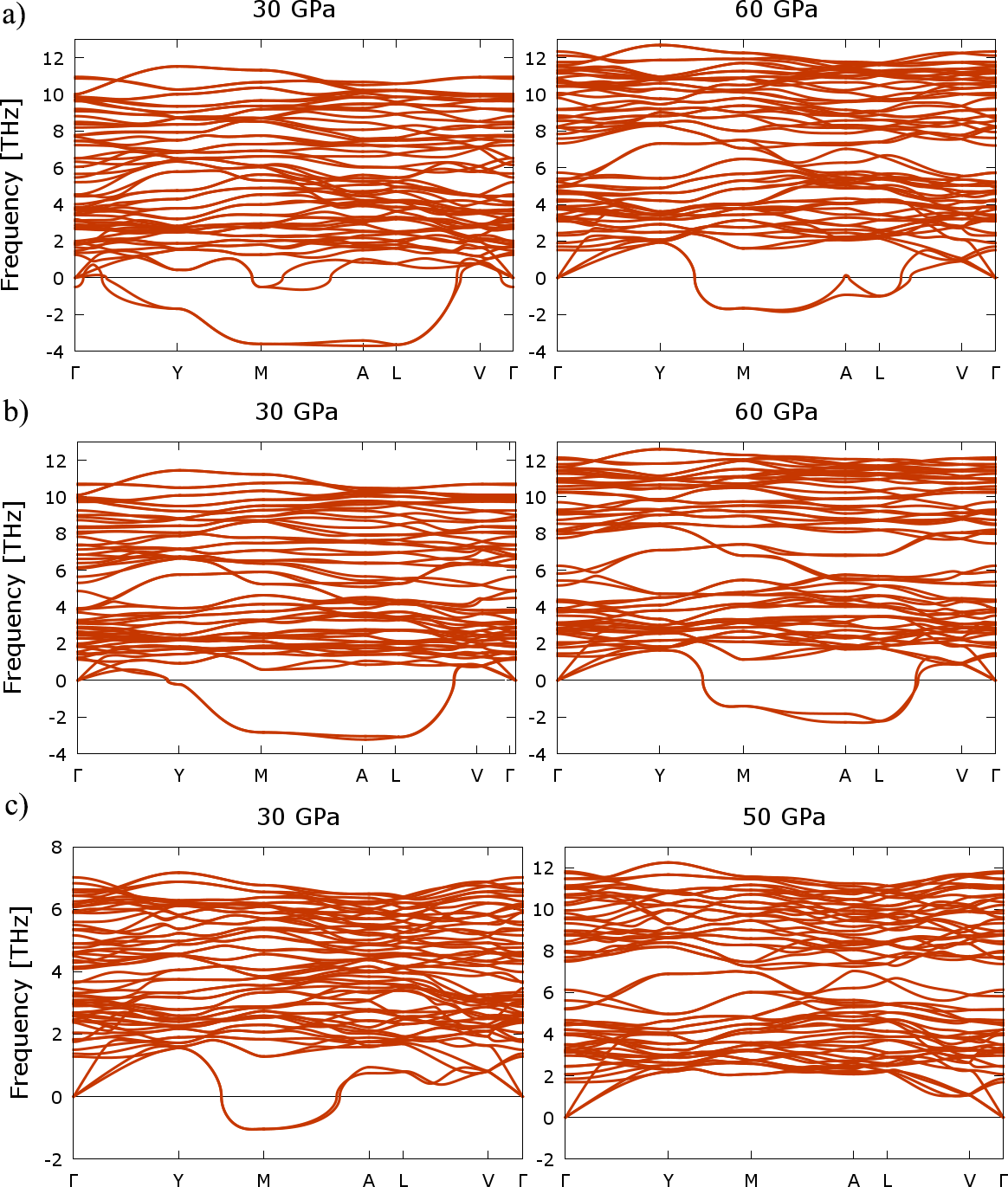}
\caption{\label{fig:phonon_Im3m}
Simulated phonon dispersion curves of the disordered bcc-like \textit{Im-3m} phases of Sb$_2$S$_3$ (a), Bi$_2$S$_3$ (b) and Sb$_2$Se$_3$ (c), and calculated at 30 GPa (left) and 50 (Sb$_2$Se$_3$) or 60 GPa (Sb$_2$S$_3$ and Bi$_2$S$_3$; right). The BZ \textbf{q}-vector description represents the \textit{C2/m} space-group, according to the symmetry of the employed cell.} 
\end{center}
\end{figure*}

As illustrated in Fig. \ref{fig:phonon_Im3m}, at 30 GPa all three disordered structures show negative modes along the dispersion curves, thus indicating that these structures are dynamically unstable at this pressure range. 

At 60 GPa the phonon dispersion curves of Sb$_2$S$_3$ and Bi$_2$S$_3$ present imaginary modes (Fig. \ref{fig:phonon_Im3m}), indicating that neither compound is likely to adopt this phase for pressures, at least until 60 GPa. We note however that the dynamical instabilities found for Sb$_2$S$_3$ and Bi$_2$S$_3$ both decrease (the negative modes shift to higher frequency values, towards positive values) with increasing pressure, suggesting that these phases could in principle become stable at pressures above 60 GPa. In this context, we must note that Efthimiopoulos \textit{et al}.,\cite{JPhysChemA.118.1713.2014} had observed a pressure-induced amorphization above 50 GPa for Bi$_2$S$_3$, however the authors were not able to identify the phase to be the disordered \textit{Im-3m} structure, even at 65 GPa. On the other hand, experimental data for Sb$_2$S$_3$, suggests that the disordered bcc-like phase exists between 28.2 and 50.2 GPa.\cite{SciRep.8.14795.2018} However, it must be noted, that experimental measurements detailed in Ref. \onlinecite{SciRep.8.14795.2018} were carried out under non-hydrostatic behaviour, due to the employed pressure-transmitting medium.

Finally, our calculations suggest that Sb$_2$Se$_3$ becomes dynamically stable already at 50 GPa; a result that is in agreement with the \textit{Im-3m} phase being observed experimentally around 50 GPa.\cite{SciRep.3.2665.2013}

To close this point, we can speculate that the stability of the disordered solid solution of sesquichalcogenides seems to be related to the size of cations and anions since the \textit{Im-3m} phase is consistently being observed at HP for sesquichalcogenides with heavier cations and anions (Sb$_2$Se$_3$, Sb$_2$Te$_3$, Bi$_2$Se$_3$, and Bi$_2$Te$_3$). It seems that the possibility of occuring such a HP phase could be related to the radii size\cite{Seeger.Wiley.2007} of Se, Te, Sb and Bi (atomic radii: $r_\textrm{Se}=117$, $r_\textrm{Te}=137$, $r_\textrm{Sb}=141$ and $r_\textrm{Bi}=182$ pm, respectively). Stemming on these values, we can infer that the solid solutions are energetically favourable in sesquichalcogenides if the atomic radii of the cation and anion differ by less than $\sim 65$ pm, or if the size ratio between them is smaller than 1.55 (case of Bi$_2$Se$_3$). It is thus likely that the disordered \textit{Im-3m} phase of Sb$_2$S$_3$ could indeed stabilize, because the radii difference between $r_\textrm{Sb}$ and $r_\textrm{S}$ is 37 pm (141-104=37 pm) and the size ratio is 1.35, and therefore within the above mentioned thresholds. However, the larger radius of Bi results in a larger radii difference (78 pm) and ratio (1.75) with respect to S, which could therefore evidence the instability of such a disordered phase for Bi$_2$S$_3$ at HP. 

Moreover and as suggested in Ref.~\onlinecite{PhysRevLett.106.145501.2011}, the atomic radii between the anion and cation tends to become approximately equal under pressure due to a higher probability of charge transfer from cation to anion. Therefore, HP inherently creates a favourable enviroment for the disordered phase due to the decrease of the difference between cation and anion atomic radii. Consequently, the transition to the Bi$_2$S$_3$ disordered solid solution could probably be induced for very high pressure values, namely when the difference between the two radii decreases below 65 pm and the ratio decreases below 1.55. %becomes approximately equal, $r_\textrm{Bi}/r_\textrm{S} \rightarrow 1$. 

\subsubsection{The Low-Pressure \textit{Pnma} Phase at High-Pressure}

\begin{figure}
\begin{center}
\includegraphics[width=8cm]{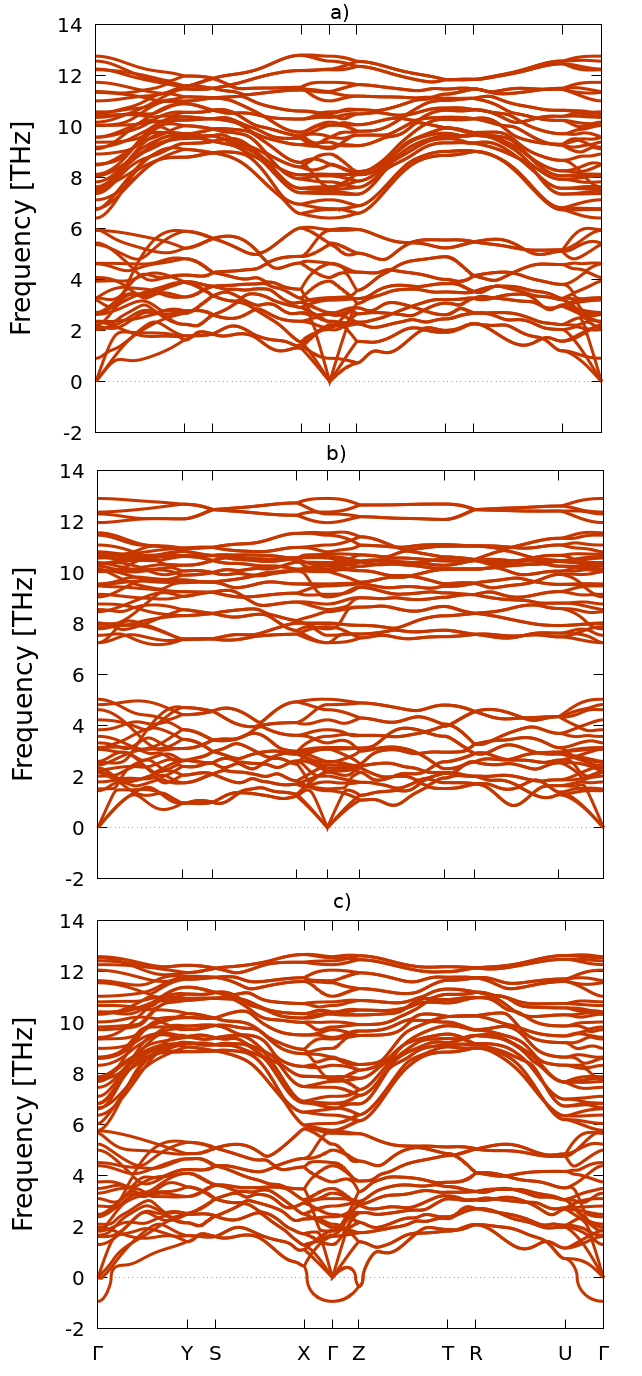}
\caption{\label{fig:phonon_Pnma}
Simulated phonon dispersion curves of the \textit{Pnma} phases of Sb$_2$S$_3$ (a), Bi$_2$S$_3$ (b) and Sb$_2$Se$_3$ (c), calculated at 50 GPa.} 
\end{center}
\end{figure}

In order to study the dynamical stability of the well known low-pressure \textit{Pnma} phase at HP, we present in Fig. \ref{fig:phonon_Pnma} the phonon dispersion curves of the respective phase for Sb$_2$S$_3$, Bi$_2$S$_3$ and Sb$_2$Se$_3$ at 50 GPa.

Curiously enough, we find that for Sb$_2$S$_3$ and Bi$_2$S$_3$ the system is still dynamically stable at 50 GPa, although thermodynamically it is not the most stable phase (Fig. \ref{fig:enthalpy}). These results, together with the dynamical instability observed for the disordered phase of Sb$_2$S$_3$ and Bi$_2$S$_3$ at 50 GPa (Fig. \ref{fig:phonon_Im3m}), and the thermodymical instability of the \textit{C2/m} and \textit{C2/c} phases, suggest that only the \textit{Pnma} structure should be observed at 50 GPa for both compounds; upto a plausible phase-transition should occur at higher pressures values. 

For Sb$_2$Se$_3$ however, we note that at 50 GPa the \textit{Pnma} structure starts presenting negative frequencies, localised around the zone-centre, therefore evidencing dynamical unstability at the same pressure range where the disordered \textit{Im-3m} phase is already dynamically stable (Fig. \ref{fig:phonon_Im3m}). Therefore, our dynamical and thermodynamical results of Sb$_2$Se$_3$, clearly suggests that at HP a transition from the \textit{Pnma} phase to the disordered \textit{Im-3m} phase is likely to occur, in good agreement with experiment.   

\subsubsection{The Low-Pressure Phases of Sb$_2$Se$_3$}

%\bigskip 
By considering the enthalpy energies of the \textit{Pnma} vs \textit{R-3m} phases of Sb$_2$Se$_3$ (Fig. \ref{fig:enthalpy}) we have shown that the \textit{R-3m} phase is energetically the most favourable phase up to $\sim$4.8 GPa, evidencing a very low energy barrier between the \textit{Pnma} phase of only 22.71 meV (per f.u.) at 0 GPa. 

\begin{figure}
\begin{center}
\includegraphics[width=8cm]{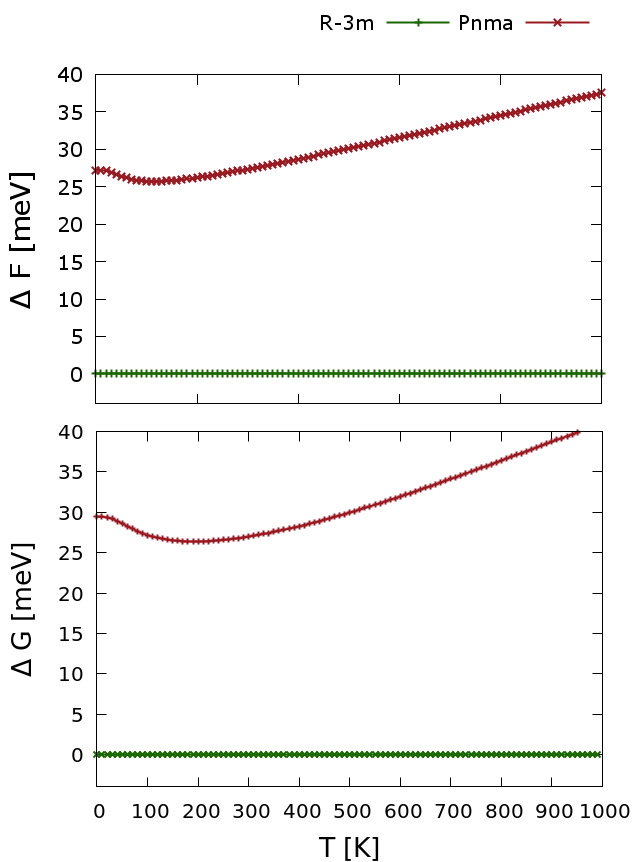}
\caption{\label{fig:free_energies}
Constant-volume Helmoltz (top) and Gibbs (bottom) free energies of the \textit{R-3m} phase (green) relative to the \textit{Pnma} phase (red) of Sb$_2$Se$_3$ as a function of temperature.} 
\end{center}
\end{figure}

In order to verify if the entropy contributions to the DFT total energies could affect the energetic stability found for \textit{R-3m} with respect to \textit{Pnma} (the experimentally observed phase), we have evaluated the constant-volume (Helmholtz) Free energy at 0 GPa (Fig. \ref{fig:free_energies}, top). The Helmholtz Free energy ($F$) is obtained by summing the lattice energy (DFT total energy) and the vibrational contributions from the population of the harmonic phonon energy levels. \cite{JChemPhys.123.204708.2005}

From Fig. \ref{fig:free_energies} (top), we may observe that at 0 GPa the \textit{R-3m} phase is the most stable phase at any temperature range, and no transitioning is observed to the \textit{Pnma} structure. At 0 K, the Free energy difference between the two phases is 27.24 meV, which is $\sim$4.53 meV higher when compared to the enthalpy energy difference at 0 K (zero-point energy). Moreover, at 300 K the energy difference between the phases has a negligible increase of 0.11 meV (27.35 meV).

Another factor that can influence in the ordering of two competing phases is the thermal expansion. Variation of the lattice volume due to thermal expansion/contraction can be introduced by the quasi-harmonic approximation (QHA), in which the thermal expansion of the lattice is obtained from the volume dependence of the phonon frequencies. The evaluation of the equilibrium volume and Gibbs free energy ($G$) at a finite temperature is thus obtained by minimising the Helmoltz Free energy for a given (constant) pressure. The theoretical background of the QHA is detailed in Refs. \onlinecite{JChemPhys.123.204708.2005} and \onlinecite{PhysRevB.91.144107.2015} and therefore will not be extended in the present work. 

Fig. \ref{fig:free_energies} (bottom) shows the difference of $G$ between the two phases of interest. One may observe that by taking into account the thermal expansion, the \textit{R-3m} phase still remains the most stable phase with respect to the \textit{Pnma} phase, with an energy difference of 29.43 meV at 0 K (very similar behaviour to that obtained from $F$). At room temperature (300 K) the energy difference between the phases decreases slightly down to 26.96 meV. In summary, at room pressure, our Free energy results show that \textit{R-3m} is always more stable than \textit{Pnma} at any temperature. Neither for $F$ nor for $G$ do the differences decrease below the k$_\mathrm{B}$T limit, and therefore the energy barrier is higher than that required for the phase transition to be spontaneous given the available thermal energy (which does not occur when the zero-point energy is not considered).

\begin{figure*}
\begin{center}
\includegraphics[width=12cm]{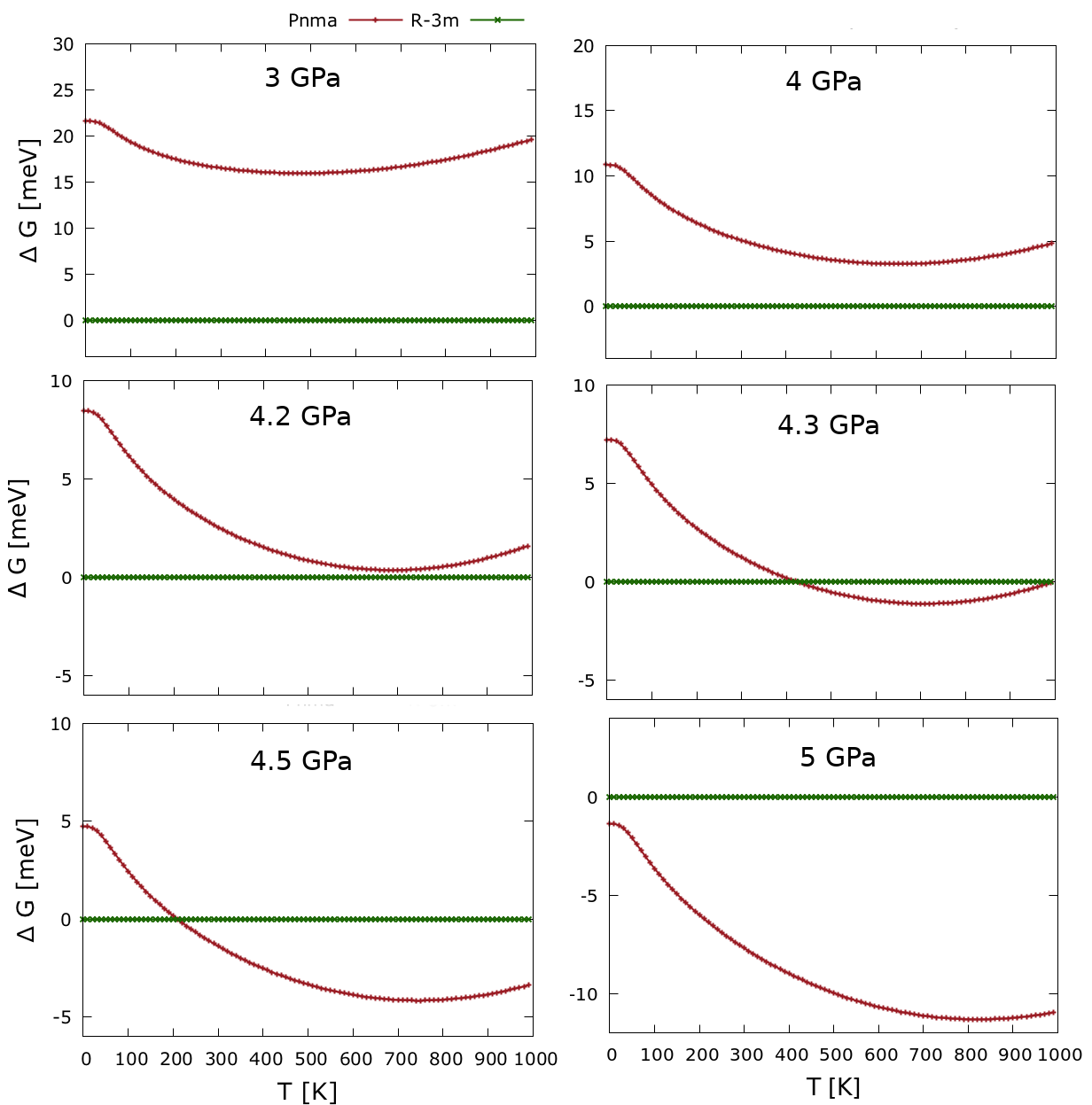}
\caption{\label{fig:free_energies_pressure}
Gibbs free energies of the \textit{R-3m} phase (green) relative to the \textit{Pnma} phase (red) of Sb$_2$Se$_3$ as a function of temperature, and for different pressure values.} 
\end{center}
\end{figure*}

We also present the $G$ differences of Sb$_2$Se$_3$ for pressure values between 0 and 5 GPa (Fig. \ref{fig:free_energies_pressure}), in order to analyse the energetic ordering between the two phases and probe if a pressure-induced phase transition could be observed as a function of temperature.  We observe that the \textit{R-3m} phase persists in the energetically stability at any temperature range up to 1000 K, for 3 and 4 GPa. However we must note that the energy differences between the two phases decreases considerably for increasing pressures and at high temperature values. At 4 GPa the lowest energy difference ($\sim$3.25 meV) between the two phases occurs between 650-700 K. Increasing the pressure slightly (4.2 GPa) results in an energy decrease of \textit{Pnma} nearly reaching the energy of \textit{R-3m} at around 400 K. In fact, at 4.3 GPa the phase transition from \textit{R-3m} to \textit{Pnma} is observed around 400 K; at 4.5 GPa the transition temperature decreases to $\sim$200 K. Finally at 5 GPa, the \textit{Pnma} phase becomes the most energetically favourable structure for the Sb$_2$Se$_3$ system. 

Based on the analysis from $G$, we conclude that the pressure-induced transition between the \textit{R-3m} to \textit{Pnma} is favoured at low pressures (between 4.2-4.4 GPa) near room temperature conditions. This conclusion is very similar to that evidenced from the enthalpy plots at 0 K (Fig. \ref{fig:enthalpy}) where the phase transition is predicted to be around 4.8 GPa. 

\begin{figure}
\begin{center}
\includegraphics[width=8cm]{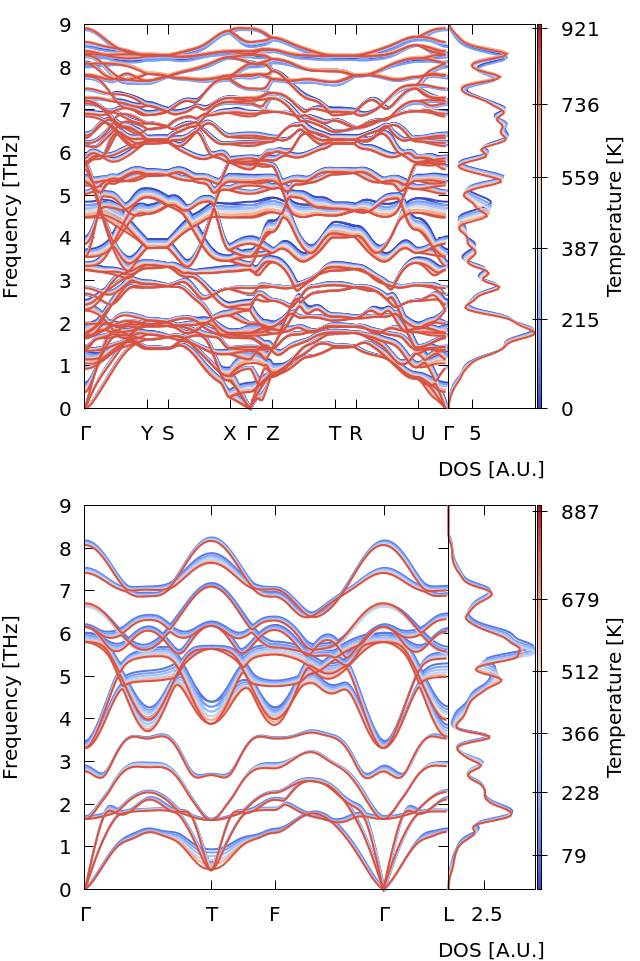}
\caption{\label{fig:qha_band}
Quasi-harmonic phonon dispersion curves for \textit{Pnma}-Sb$_2$Se$_3$ (top) and \textit{R-3m}-Sb$_2$Se$_3$ (bottom). The color gradient runs from blue (low T) to red (high T) across the temperatures associated with the volume expansions considered in our calculations. } 
\end{center}
\end{figure}

After confirming that in fact the \textit{R-3m} phase is thermodynamically more stable than the \textit{Pnma} phase at 0 GPa, at any temperature range, we probe the dynamical stability of the \textit{R-3m} compound in order to assess whether this structure could potentially be synthesized for Sb$_2$Se$_3$ at/or close to ambient conditions.

For this purpose, we have evaluated the phonon band dispersion and density of states (DoS) of the \textit{R-3m} phase phase at 0 GPa, for different temperature values (Fig. \ref{fig:qha_band}). We also present the phonon band structure and DoS for the \textit{Pnma} phase for sake of comparison. Our results show that there are no imaginary frequencies throughout the whole of the Brillouin-zone, thus indicating that both phases to be dynamically stable under ambient conditions (0 GPa and room temperature) and confirming that, as implied by the energetics comparison, both phases could potentially coexist. 

In this context, we must note that our results confirm a recent theoretical work performed on the \textit{R-3m} Sb$_2$Se$_3$ phase reporting the phonon dispersion curves, and confirming dynamical stability of this phase at 0 GPA and 0 K.\cite{PhysRevB.97.075147.2018} In this work, the formation energies of Sb$_2$Se$_3$ were also computed evidencing the \textit{R-3m} phase to be energetically stable. Moreover, \textit{ab initio} molecular dynamics\cite{PhysRevB.97.075147.2018} confirm that the \textit{R-3m} phase remains unchanged at finite temperature (300 K), once again favouring our presented results.

\subsection{Mechanical Stability}
\label{subsec:elastic}

Mechanical or elastic stability is the third condition that should be satisfied for a system to be potentially synthesized. Such a study is carried out by probing if the elastic constants obey the Born stability criteria when the solid is submitted to homogeneous deformations.\cite{ProcCambridgePhilosSoc36.160.1940, JPhysCondensMatter.9.8579.1997, PhysRevB.90.224101.2014} 

We therefore evaluate the mechanical stability of the \textit{Pnma} and \textit{R-3m} phases of Sb$_2$Se$_3$ by calculating and comparing the elastic tensors between the two low-pressure phases (Tab. \ref{table:elastic_cts}).

\begin{table*}
\begin{center}
\caption{ \label{table:elastic_cts} 
Calculated elastic constants c$_{ij}$ (GPa) of the \textit{Pnma} and \textit{R-3m} phases of Sb$_2$Se$_3$ at 0 GPa.}
\begin{tabular}{| c| c|c|c|c|c|c|c|c|c |}\hlinewd{1pt}
  & c$_{11}$  & c$_{22}$ & c$_{33}$ & c$_{12}$ & c$_{13}$ & c$_{23}$ & c$_{44}$ & c$_{55}$ & c$_{66}$\\ \hline
\textbf{\textit{Pnma}} & 30.92  & 81.65 & 55.20 &  17.32 &  15.10 & 26.89 & 17.83 & 25.21 & 7.69 \\
\hlinewd{1pt}
\end{tabular}
%\end{center}
%\end{table*}
%\begin{table*}
%\begin{center}
\begin{tabular}{| c|c|c|c|c|c|c|c |}\hlinewd{1pt}
& c$_{11}$ = c$_{22}$ & c$_{33}$ & c$_{12}$ & c$_{13}$ = c$_{23}$ & c$_{15}$ = -c$_{25}$ = c$_{46}$  & c$_{44}$ = c$_{55}$ & c$_{66}$\\\hline
\textbf{\textit{R-3m}}  & 90.81  & 40.21 & 25.85 &  21.61 &  -12.00 &  25.10 & 32.48\\ 
\hlinewd{1pt}
\end{tabular}
\end{center}
\end{table*}

\begin{table}
\begin{center}
\caption{ \label{table:elastic_prop} 
Calculated elastic properties of the \textit{Pnma} and \textit{R-3m} phases in Sb$_2$Se$_3$ at 0 GPa as obtained within the Voigt approximation using the ELATE analysis tool: bulk modulus, B$_0$, Young modulus, $E$, shear modulus, $G$ and Poisson’s ratio, $\upsilon$.}
\begin{tabular}{| c| c|c|c|c|}\hlinewd{1pt}
  & \textbf{B$_0$ (GPa)} &	\textbf{$E$ (GPa)} &	\textbf{$G$ (GPa)} &  $\mathbf{\upsilon}$\\ \hline
\textbf{\textit{Pnma}} & 31.82	& 44.10	& 17.38 &	0.27 \\ \hline
\textbf{\textit{R-3m}}  & 40.00	& 65.56 &	26.72 &	0.23\\ 
\hlinewd{1pt}
\end{tabular}
\end{center}
\end{table}

 To confirm the accuracy of our calculated elastic constants, we have computed the linear compressibility of both phases at 0 GPa using the ELATE analysis tools. For both phases, only directions corresponding to positive linear compressibilities were obtained, indicating both phases to be mechanically stable under ambient conditions. In the case of the \textit{R-3m} phase, we have obtained linear compressibilities between $\beta_\textrm{min}$ = 4.9 TPa$^{-1}$ (hexagonal a-axis) and $\beta_\textrm{max}$ = 19.5 TPa$^{-1}$ (hexagonal c-axis) with an anisotropy value of 3.95. For the \textit{Pnma} phase, the compressibilities fall between $\beta_\textrm{min}$ = 3.7 TPa$^{-1}$ (b-axis) and $\beta_\textrm{max}$ = 25.7 TPa$^{-1}$ (a-axis) with an anisotropy of 6.87. These values are of the same order as the experimental axial compressibilities of the \textit{Pnma} phase of Sb$_2$Se$_3$ ($\beta_\textrm{a}$ = 15.2 TPa$^{-1}$, $\beta_\textrm{b}$ = 3.9 TPa$^{-1}$, $\beta_\textrm{c}$ = 8.3 TPa$^{-1}$).\cite{JPhysChemA.118.1713.2014} We must stress that the bulk modulus of the \textit{Pnma} phase of Sb$_2$Se$_3$ calculated from the elastic-constant tensor (31.8 GPa), and summarized in Tab. \ref{table:elastic_prop} together with other elastic moduli, is similar to that obtained from the Birch-Murnaghan fit (c.f. Tab. \ref{table:pnma_param}), as expected, thus demonstrating the adequate convergence criteria employed throughout the present calculations.
The calculated elastic constants in Tab. \ref{table:elastic_cts} fulfill the necessary and sufficient Born criteria for the mechanical stability of orthorhombic (Eq. \ref{eq:pnma}) and rhombohedral (Eq. \ref{eq:r-3m}) systems, respectively.\cite{PhysRevB.90.224101.2014} Therefore, our calculated elastic constants indicate that both the \textit{R-3m} and \textit{Pnma} phases of Sb$_2$Se$_3$ are mechanically stable under ambient conditions.

\begin{eqnarray}
&& c_{11},c_{44},c_{55},c_{66}>0; \nonumber\\
&& c_{11}c_{22}>c_{12}^2; \nonumber\\ 
&& c_{11}c_{22}c_{33}+2c_{12}c_{13}c_{23}\nonumber\\
&&-c_{11}c_{23}^2-c_{22}c_{13}^2-c_{33}c_{12}^2>0 \label{eq:pnma}
\\\nonumber\\
&& c_{11}>|c_{12}|; c_{44}>0;\nonumber\\
&& c_{13}^2<\frac{1}{2}c_{33}(c_{11}+c_{12}); \nonumber\\
&& c_{14}^2<\frac{1}{2}c_{44}(c_{11}-c_{12})=c_{44}c_{66} 
\label{eq:r-3m}
\end{eqnarray}

We note that elastic tensors of Sb$_2$Se$_3$ have been previously calculated,\cite{SolidStateSci.14.1211.2012} although as for the bulk modulus, respective components were overestimated as well. We believe a reason for the disagreement could be due to the low cut-off energy used in those calculations.

%%%%%%%%%%%%%%%%%%%%%%%%%%%%%%%%%%%%
\subsection{Lattice Parameters, Infrared and Raman Spectra of the \textit{R-3m} phase of Sb$_2$Se$_3$}
%%%%%%%%%%%%%%%%%%%%%%%%%%%%%%%%%%%%

Our calculations show that the \textit{R-3m} phase of Sb$_2$Se$_3$ is energetically competitive with the \textit{Pnma} phase and is both mechanically and dynamically stable, all of which suggest this phase should be a ground-state structure under ambient conditions.
 
The inconsistency found between the theoretical and experimental data regarding Sb$_2$Se$_3$, can be based on the possibility that the \textit{Pnma} phase forms faster than the \textit{R-3m} phase, under the usual synthesis conditions. In fact, the \textit{R-3m} phase has not been proposed on the pressure/temperature phase diagram prepared by Pfeiffer \textit{et al.},\cite{pfeiffer.PhDthesis.2009} although this study did not attempt to vary the synthesis conditions at close to ambient pressure, as the present calculations suggest. In any case, it is noteworthy of mentioning that our calculations indicate the \textit{R-3m} phase of Sb$_2$Se$_3$ is energetically competitive with the \textit{Pnma} phase at close to ambient conditions. This raises the possibility that the \textit{R-3m} phase could potentially be prepared under slightly non-equilibrium conditions. We must note that Bera \textit{et al}.,\cite{PhysRevLett.110.107401.2013} have claimed of having observed such a phase at room temperature, although such an observation has not been confirmed by any other experimental group up untill now.

In order to assist with the possible experimental synthesis of this phase, we provide the calculated lattice parameters and atomic positions of our optimised zero-pressure \textit{R-3m} structure in Table \ref{table:lattice_r-3m}. 

\begin{table}
\begin{center}
\caption{ \label{table:lattice_r-3m} 
Predicted lattice constants and atomic positions for the hexagonal unit cell of the \textit{R-3m} phase in Sb$_2$Se$_3$ at 0 GPa.}
\begin{tabular}{| c| c|c|c|c|}\hlinewd{1pt}
   \textbf{a$_0$ (\AA)}	& \textbf{c$_0$ (\AA)}	& \textbf{V$_0$ (\AA$^3$)} &	\textbf{B$_0$ (GPa)} &	\textbf{B$_0'$}\\ \hline
 4.01 &	28.16 &	392.16	& 50.56	& 4.16 \\ 
\hlinewd{1pt}
\end{tabular}
\begin{tabular}{| c|c|c|c|c|c|}\hlinewd{1pt}
& \textbf{Site}	& \textbf{Sym.}	& \textbf{x}	 & \textbf{y}	& \textbf{z}\\\hline
\textbf{Sb$_1$}	 & 6c	& 3m &	0.00000	& 0.00000	& 0.60082 \\ \hline
\textbf{Se$_1$}	& 3a	 & -3m	& 0.00000	& 0.00000	& 0.00000\\  \hline
\textbf{Se$_2$}	& 6c & 	3m	& 0.00000	& 0.00000	& 0.78792\\ 
\hlinewd{1pt}
\end{tabular}
\end{center}
\end{table}

\begin{figure}
\begin{center}
\includegraphics[width=8cm]{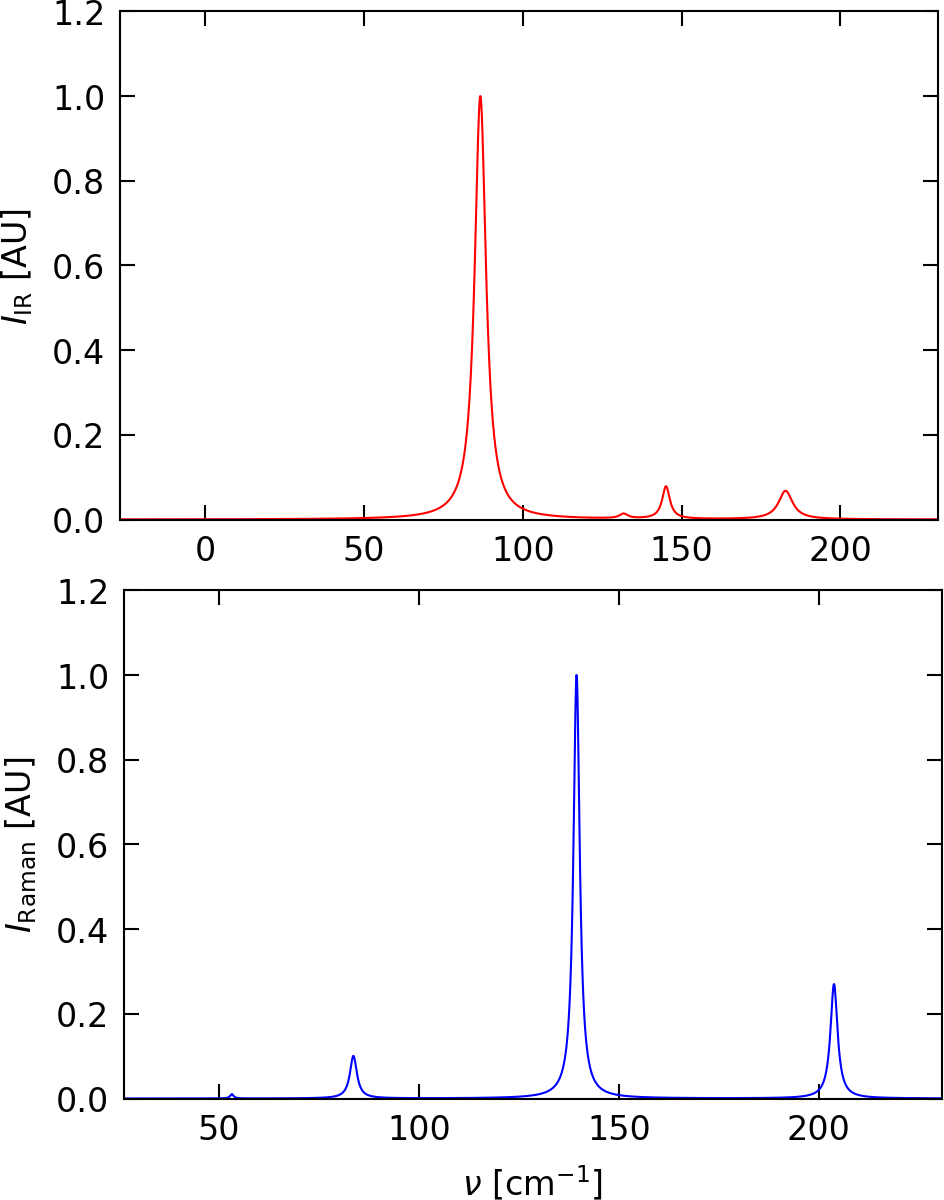}
\caption{\label{fig:raman_Sb2Se3}
Simulated infrared (IR; top) and Raman (bottom) spectra of the \textit{R-3m} phase of Sb$_2$Se$_3$ at 0 GPa. The spectral lines have been broadened with the calculated intrinsic mode linewidths at 300 K.} 
\end{center}
\end{figure}

We have also computed the IR and Raman spectra to aid in the identification of the spectral signatures that should distinguish the \textit{R-3m} from the \textit{Pnma} phase (Fig. \ref{fig:raman_Sb2Se3}).\cite{SciRep.3.2665.2013} 
 
 The frequencies, irreducible representations and IR/Raman intensities associated with each of the zone-centre ($\Gamma$-point) vibrational modes are listed in Table \ref{table:ir-raman_r-3m}.
 
\begin{table}
\begin{center}
\caption{ \label{table:ir-raman_r-3m} 
Calculated vibrational modes of the \textit{R-3m} phase of Sb$_2$Se$_3$ at 0 GPa. The three acoustic IR-active modes are formed by the irreducible representations of $\Gamma_\textrm{acoustic}$ = A$_\textrm{2u}$ + E$_\textrm{u}$ and the remaining 12 optical modes are $\Gamma_\textrm{optical}$ = 2E$_\textrm{g}$ (Raman) + 2A$_\textrm{1g}$ (Raman) + 2E$_\textrm{u}$ (IR) + 2A$_\textrm{2u}$ (IR).}
\begin{tabular}{| c| c|c|c|}\hlinewd{1pt}
 \textbf{ Irr.eps.}	& \textbf{Frequency}	& \textbf{Raman Intensity} 	& \textbf{Intensity} \\
         	& \textbf{(cm$^{-1}$)}	& \textbf{(10$^5$ \AA$^4$ amu$^{-1}$)}	& \textbf{(e$^2$ amu$^{-1}$)}\\ \hline
 \textbf{E$_\textrm{g}$} &	53.3 & 	0.02 &	Inactive\\ \hline
\textbf{A$_\textrm{1g}$} &	83.7	 & 0.37	& Inactive\\ \hline
\textbf{E$_\textrm{u}$} & 	86.6	 & Inactive	& 3.47\\ \hline
\textbf{E$_\textrm{u}$}& 	131.7 &	Inactive &	0.03\\ \hline
\textbf{E$_\textrm{g}$} & 	139.4 & 3.32	 & Inactive\\ \hline
\textbf{A$_\textrm{2u}$} &	145.1	& Inactive	& 0.18\\ \hline
\textbf{A$_\textrm{2u}$} &	182.8	& Inactive & 	0.29\\ \hline
\textbf{A$_\textrm{1g}$} &	203.8	& 1.02	& Inactive\\
\hlinewd{1pt}
\end{tabular}
\end{center}
\end{table} 
 
The inversion symmetry in the \textit{R-3m} structure leads to mutual exclusion between the IR and Raman activity of the modes, with each spectrum being characterised by four bands.\cite{PhysStatusSolidiB.250.669.2013} The most intense Raman band occurs around 139 cm$^{-1}$ (E$_\textrm{g}$), while a second prominent feature is predicted at $\sim$204 cm$^{-1}$ (A$_\textrm{1g}$). The frequency of this A$_\textrm{1g}$ mode is higher than in Bi$_2$Se$_3$ but lower than in In$_2$Se$_3$, as expected from the difference in mass between In, Sb and Bi.\cite{PhysRevB.84.184110.2011,InorgChem.57.8241.2018} Lower-frequency E$_\textrm{g}$ and A$_\textrm{1g}$ modes with much lower intensities are also found around 53 and 84 cm$^{-1}$, respectively, which are again slightly higher than the corresponding frequencies calculated for Bi$_2$Se$_3$.\cite{PhysRevB.84.184110.2011} There are four IR-active modes, two with E$_\textrm{1u}$ symmetry (87 and 132 cm$^{-1}$) and two with A$_\textrm{2u}$ bands (145 and 183 cm$^{-1}$). Of these, the 87 cm$^{-1}$ mode is the most prominent in the spectrum, while the second E$_\textrm{u}$ mode at 132 cm$^{-1}$ is very weak. The two A$_\textrm{2u}$ bands have comparable, moderate intensities and form a pair of smaller features at higher frequencies. As expected given the mass difference, the IR-active modes in Sb$_2$Se$_3$ again have slightly higher frequencies than those calculated for Bi$_2$Se$_3$.\cite{PhysRevB.84.184110.2011}

%%%%%%%%%%%%%%%%%%%%%%%%%%%%%%%%%%%%
\section{Conclusions}
%%%%%%%%%%%%%%%%%%%%%%%%%%%%%%%%%%%%

In summary, we have carried out a comprehensive set of calculations to investigate the stability of five possible phases, \textit{viz}. \textit{Pnma}, \textit{R-3m}, \textit{C2/m}, \textit{C2/c} and disordered \textit{Im-3m} of the Sb$_2$S$_3$, Bi$_2$S$_3$ and Sb$_2$Se$_3$ sesquichalcogenides under hydrostatic pressures up to 60 GPa.

For the three coumpounds we find that the monoclinic \textit{C2/m} and \textit{C2/c} phases are energetically less favourable throughout the studied pressure range and are not expected to be observed at HP under hydrostatic conditions. On the other hand, the disordered bcc-like \textit{Im-3m} phase is predicted to be the most energetically stable phase of the three compounds at HP. However, calculated phonon dispersion curves indicate that such a structural phase remains dynamically unstable up to at least 60 GPa for Sb$_2$S$_3$ and Bi$_2$S$_3$. Moreover, the \textit{Pnma} phase is stable for the two compounds at HP, even up to 50 GPa; therefore, we do not expect to observe the disordered \textit{Im-3m} phase in either of these two compounds below 50 GPa. This conclusion agrees with results from Ref. \onlinecite{JPhysChemA.118.1713.2014}, where the disordered phase has not been observed for Bi$_2$S$_3$ up to 65 GPa. However, our results disagree for Sb$_2$S$_3$ with what was reported in Ref. \onlinecite{SciRep.6.24246.2016}, regarding the observation of multiple high-pressure phases. This effect could have been caused by the use of the specific pressure-transmitting medium resulting in non-hydrostatic behaviour during the experimental measurements.

For Sb$_2$Se$_3$ our calculations predict a transition to occur from the \textit{Pnma} to the disordered \textit{Im-3m} phase above 21 GPa. 
Unlike the former two compounds, at 50 GPa, the \textit{Pnma} phase begins to evidence negative phonons at the $\Gamma$-point, thus indicating dynamical instability at HP; whereas the disordered \textit{Im-3m} phase stabilizes at this pressure range. Our calculations therefore support the conclusion that a phase transition occurs for Sb$_2$Se$_3$ from the \textit{Pnma} to the \textit{Im-3m} phase at HP, in good agreement with experimental findings. \cite{SciRep.3.2665.2013}

By probing the low-pressure regions, we find that the \textit{Pnma} phase is the most stable phase for Bi$_2$S$_3$ and Sb$_2$S$_3$, in good agreement with experiments. However, and unexpectedly, for Sb$_2$Se$_3$, it is the \textit{R-3m} phase that possess lower energy at 0 GPa, being surpassed by the \textit{Pnma} at a moderate pressure range, slightly below 5 GPa and around room-temperature conditions. Since the \textit{Pnma} phase is experimentally obtained for Sb$_2$Se$_3$, we suggest that the orthorhombic phase is stabilized by thermal energy at room temperature, which suggests the possibility of synthesizing the \textit{R-3m} phase under optimized conditions. We would expect the trigonal phase to show topological insulating properties under ambient conditions, which would likely make such an undertaking highly worthwhile. We therefore provide theoretical lattice parameters and atomic positions along with reference IR and Raman spectra to aid future experiments to identify and characterize this phase. We hope that this work will stimulate further investigation of the sesquichalcogenides at high pressure, especially to the lesser-known As analogues of the compounds examined in this work.

%\section{Appendix}
%
%%%%%%%%%%%%%%%%%%%%%%%%%%%%%%%%%%%%%%%%%%%%
%\subsection{Temperature Dependant Properties of the \textit{Pnma} and \textit{R-3m} Phases of Sb$_2$Se$_3$}
%%%%%%%%%%%%%%%%%%%%%%%%%%%%%%%%%%%%%%%%%%%%

%%%%%%%%%%%%%%%%%%%%%%%%%%%%%%%%%%%%
\section{Acknowledgements}
%%%%%%%%%%%%%%%%%%%%%%%%%%%%%%%%%%%%

This research was supported by the Spanish Ministerio de Ciencia, Innovaci\'{o}n y Universidades under the projects MAT2016-75586-C4-1-P/2-P/3-P and RED2018-102612-T (MALTA-Consolider-Team network), by the Generalitat Valenciana under project PROMETEO/2018/123 (EFIMAT); and by the European Union Horizon 2020 research and innovation programme under Marie Sklodowska-Curie grant agreement No. 785789-COMEX. JMS is grateful to the University of Manchester for the support of a Presidential Fellowship. The authors acknowledge the use of the MALTA Computing Cluster at the University of Oviedo and the computer resources at MareNostrum with technical support provided by the Barcelona Supercomputing Center (QCM-2018-3-0032).

\clearpage

\clearpage 
\bibliographystyle{unsrt}
%%%\bibliographystyle{abbrv} 
%\bibliography{biblo}

\begin{thebibliography}{10}

\bibitem{Science.325.178.2009}
Y.~L. Chen, J.~G. Analytis, J.-H. Chu, Z.~K. Liu, S.-K. Mo, X.~L. Qi, H.~J.
  Zhang, D.~H. Lu, X.~Dai, Z.~Fang, S.~C. Zhang, I.~R. Fisher, Z.~Hussain, and
  Z.-X. Shen.
\newblock Experimental realization of a three-dimensional topological
  insulator, {B}i$_2${T}e$_3$.
\newblock {\em Science}, 325(5937):178, 2009.

\bibitem{NatPhys.5.438.2009}
H.~Zhang, C.-X. Liu, X.-L. Qi, X.~Dai, Z.~Fang, and S.-C. Zhang.
\newblock Topological insulators in {B}i$_2${S}e$_3$, {B}i$_2${T}e$_3$ and
  {S}b$_2${T}e$_3$ with a single dirac cone on the surface.
\newblock {\em Nat. Phys.}, 5:438, 2009.

\bibitem{RevModPhys.282.3045.2010}
M.~Z. Hasan and C.~L. Kane.
\newblock Colloquium: Topological insulators.
\newblock {\em Rev. Mod. Phys.}, 82:3045, Nov 2010.

\bibitem{JPhysChemSolids.2.240.1957}
J.~Black, E.~M. Conwell, L.~Seigle, and C.~W. Spencer.
\newblock Electrical and optical properties of some
  {M}$_2$$^{V-B}${N}$_3$$^{VI-B}$ semiconductors.
\newblock {\em J. Phys. Chem. Solids}, 2(3):240, 1957.

\bibitem{PhysRevB.87.205125.2013}
M.~R. Filip, C.~E. Patrick, and F.~Giustino.
\newblock $gw$ quasiparticle band structures of stibnite, antimonselite,
  bismuthinite, and guanajuatite.
\newblock {\em Phys. Rev. B}, 87:205125, May 2013.

\bibitem{JPhysChemLett.1.1524.2010}
S.-J. Sa~Moon, Y.~Itzhaik, J.-H. Yum, S.~M. Zakeeruddin, G.~Hodes, and
  M.~Gr\"{at}zel.
\newblock {S}b$_2${S}$_3$-based mesoscopic solar cell using an organic hole
  conductor.
\newblock {\em J. Phys. Chem. Lett.}, 1:1524, 2010.

\bibitem{AdvFunctMater.21.4663.2011}
C.~E. Patrick and F.~Giustino.
\newblock Structural and electronic properties of semiconductor-sensitized
  solar-cell interfaces.
\newblock {\em Adv. Funct. Mater.}, 21(24):4663, 2011.

\bibitem{NatPhotonics.9.409.2015}
Y.~Zhou, L.~Wang, S.~Chen, S.~Qin, X.~Liu, J.~Chen, D.‐J. Xue, M.~Luo,
  Y.~Cao, Y.~Cheng, E.~Sargent, and J.~Tang.
\newblock Thin-film {S}b$_2${S}e$_3$ photovoltaics with oriented
  one-dimensional ribbons and benign grain boundaries.
\newblock {\em Nat. Photonics}, 9:409, 05 2015.

\bibitem{NatEnergy.2.17046.2017}
L.~Wang, D.~Li, K.~Li, C.~Chen, H.-X. Deng, L.~Gao, Y.~Zhao, F.~Jiang, L.~Li,
  F.~Huang, Y.~He, H.~Song, G.~Niu, and J.~Tang.
\newblock Stable 6\%-efficient {S}b$_2${S}e$_3$ solar cells with a {Z}n{O}
  buffer layer.
\newblock {\em Nat. Energy}, 2:17046, 03 2017.

\bibitem{NatMater.5.118.2006}
O.~Rabin, J.~Perez, J.~Grimm, G.~Wojtkiewicz, and R.~Weissleder.
\newblock An {X}-ray computed tomography imaging agent based on
  long-circulating bismuth sulphide nanoparticles.
\newblock {\em Nat. materials}, 5:118, 02 2006.

\bibitem{PhysChemB.110.21408.2006}
K.~Yao, Z.~Zhang, X.~Liang, Q.~Chen, L.-M. Peng, and Y.~Yu.
\newblock Effect of {H}$_2$ on the electrical transport properties of single
  {B}i$_2${S}$_3$ nanowires.
\newblock {\em J. Phys. Chem. B}, 110:21408, 11 2006.

\bibitem{NanoLett.9.1482.2009}
L.~Cademartiri, F.~Scotognella, P.~O'Brien, B.~Lotsch, J.~Thomson, S.~Petrov,
  N.~Kherani, and G.~Ozin.
\newblock Cross-linking {B}i$_2${S}$_3$ ultrathin nanowires: A platform for
  nanostructure formation and biomolecule detection.
\newblock {\em Nano Lett.}, 9:1482, 05 2009.

\bibitem{arxiv.1410.2783.2014}
W.~A. Crichton, F.~L.~M. Bernal, J.~Guignard, M.~Hanfland, and S.~Margadonna.
\newblock Observation of the {S}b$_2${S}$_3$-type
  post-post-{G}d{F}e{O}$_3$-perovskite: A model structure for high density
  {ABX}$_3$ and {A}$_2${X}$_3$ phases, 2014.

\bibitem{MineralMRCag.80.659.2016}
W.~A. Crichton, F.~L. Bernal, J.~Guignard, M.~Hanfland, and S.~Margadonna.
\newblock {Observation of {S}b$_2${S}$_3$-type post-post-perovskite in
  {N}a{F}e{F}$_3$. Implications for {ABX}$_3$ and {A}$_2${X}$_3$ systems at
  ultrahigh pressure}.
\newblock {\em Mineral. Mag.}, 80(4):659, 2016.

\bibitem{SciRep.3.2665.2013}
I.~Efthimiopoulos, J.~Zhang, M.~Kucway, C.~Park, R.~Ewing, and Y.~Wang.
\newblock {S}b$_{2}${S}e$_{3}$ under pressure.
\newblock {\em Sci. Rep.}, 3:2665, 09 2013.

\bibitem{JPhysChemA.118.1713.2014}
I.~Efthimiopoulos, J.~Kemichick, X.~Zhou, S.~Khare, D.~Ikuta, and Y.~Wang.
\newblock High-pressure studies of {B}i$_2${S}$_3$.
\newblock {\em J. Phys. Chem. A}, 118, 02 2014.

\bibitem{JPhysCondMatter.28.015602.2016}
Y.~A. Sorb, V.~Rajaji, P.~S. Malavi, U.~Subbarao, P.~Halappa, S.~C. Peter,
  S.~Karmakar, and C.~Narayana.
\newblock Pressure-induced electronic topological transition in
  {S}b$_2${S}$_3$.
\newblock {\em J. Phys. Condens. Matter}, 28(1):015602, dec 2015.

\bibitem{JPhysChemC.120.10547.2016}
J.~Iba\~nez, J.~A. Sans, C.~Popescu, J.~L\'{o}pez-Vidrier, J.~J.
  Elvira-Betanzos, V.~P. Cuenca-Gotor, O.~Gomis, F.~J. Manj\'{o}n,
  P.~Rodr\'{i}guez-Hern\'{a}ndez, and A.~Mu\~{n}oz.
\newblock Structural, vibrational, and electronic study of {S}b$_2${S}$_3$ at
  high pressure.
\newblock {\em J. Phys. Chem. C}, 120(19):10547, 2016.

\bibitem{JAlloysCompd.688.329.2016}
C.~Li, J.~Zhao, Q.~Hu, Z.~Liu, Z.~Yu, and H.~Yan.
\newblock Crystal structure and transporting properties of {B}i$_2${S}$_3$
  under high pressure: Experimental and theoretical studies.
\newblock {\em J. Alloys Compd.}, 688:329, 2016.

\bibitem{SciRep.4.6679.2014}
P.~Kong, F.~Sun, L.~Xing, J.~Zhu, S.~Zhang, W.~Li, X.~Wang, S.~Feng, X.~Yu,
  J.~Zhu, R.~C. Yu, W.~Yang, G.~Shen, Y.~Zhao, R.~Ahuja, H.~Mao, and C.~Jin.
\newblock Superconductivity in strong spin orbital coupling compound
  {S}b$_2${S}e$_3$.
\newblock {\em Sci. Rep.}, 4:6679, 11 2014.

\bibitem{PhysRevB.97.235306.2018}
S.~Das, A.~Sirohi, G.~Kumar~Gupta, S.~Kamboj, A.~Vasdev, S.~Gayen,
  P.~Guptasarma, T.~Das, and G.~Sheet.
\newblock Discovery of highly spin-polarized conducting surface states in the
  strong spin-orbit coupling semiconductor {S}b$_{2}${S}e$_{3}$.
\newblock {\em Phys. Rev. B}, 97:235306, Jun 2018.

\bibitem{SciRep.6.24246.2016}
I.~Efthimiopoulos, C.~Buchan, and Y.~Wang.
\newblock Structural properties of {S}b$_2${S}$_3$ under pressure: Evidence of
  an electronic topological transition.
\newblock {\em Sci. Rep.}, 6:24246, 04 2016.

\bibitem{PhysRevB.97.024103.2018}
L.~Dai, K.~Liu, H.~Li, L.~Wu, H.~Hu, Y.~Zhuang, Y.~Linfei, C.~Pu, and P.~Liu.
\newblock Pressure-induced irreversible metallization accompanying the phase
  transitions in {S}b$_2${S}$_3$.
\newblock {\em Phys. Rev. B}, 97:024103, 01 2018.

\bibitem{SciRep.8.14795.2018}
Y.~Wang, M.~Yanmei, G.~Liu, J.~Wang, Y.~Li, Q.~Li, J.~Zhang, Y.~Ma, and G.~Zou.
\newblock Experimental observation of the high pressure induced substitutional
  solid solution and phase transformation in {S}b$_2${S}$_3$.
\newblock {\em Sci. Rep.}, 8:14795, 12 2018.

\bibitem{PhysStatusSolidiB.250.669.2013}
F.~J. Manj\'{o}n, R.~Vilaplana, O.~Gomis, E.~P\'{e}rez-Gonz\'{a}lez,
  D.~Santamar\'{i}a-P\'{e}rez, V.~Mar\'{i}n-Borr\'{a}s, A.~Segura,
  J.~Gonz\'{a}lez, P.~Rodr\'{i}guez-Hern\'{a}ndez, A.~Mu\~{n}oz, C.~Drasar,
  V.~Kucek, and V.~Mu\~{n}oz Sanjos\'{e}.
\newblock High-pressure studies of topological insulators {B}i$_2${S}e$_3$,
  {B}i$_2${T}e$_3$, and {S}b$_2${T}e$_3$.
\newblock {\em Phys. Status Solidi B}, 250(4):669, 2013.

\bibitem{PhysRevB.84.245105.2011}
W.~Liu, X.~Peng, C.~Tang, L.~Sun, K.~Zhang, and J.~Zhong.
\newblock Anisotropic interactions and strain-induced topological phase
  transition in {S}b$_{2}${S}e$_{3}$ and {B}i$_{2}${S}e$_{3}$.
\newblock {\em Phys. Rev. B}, 84:245105, Dec 2011.

\bibitem{PhysRevB.89.035101.2014}
W.~Li, X.~Wei, J.-X. Zhu, C.~Ting, and Y.~Chen.
\newblock Pressure induced topological quantum phase transition in
  {S}b$_{2}${S}e$_{3}$.
\newblock {\em Phys. Rev. B}, 89:035101, 04 2013.

\bibitem{PhysRevLett.110.107401.2013}
A.~Bera, K.~Pal, D.~V.~S. Muthu, S.~Sen, P.~Guptasarma, U.~V. Waghmare, and
  A.~K. Sood.
\newblock Sharp raman anomalies and broken adiabaticity at a pressure induced
  transition from band to topological insulator in {S}b$_{2}${S}e$_{3}$.
\newblock {\em Phys. Rev. Lett.}, 110:107401, Mar 2013.

\bibitem{PhysRevB.97.075147.2018}
G.~Cao, H.~Liu, J.~Liang, L.~Cheng, D.~Fan, and Z.~Zhang.
\newblock Rhombohedral {S}b$_{2}${S}e$_{3}$ as an intrinsic topological
  insulator due to strong van der waals interlayer coupling.
\newblock {\em Phys. Rev. B}, 97:075147, Feb 2018.

\bibitem{JPhysChemLett.9.5785.2018}
G.~Liu, Z.~Yu, H.~Liu, S.~A.~T. Redfern, X.~Feng, X.~Li, Y.~Yuan, K.~Yang,
  N.~Hirao, S.~I. Kawaguchi, X.~Li, L.~Wang, and Y.~Ma.
\newblock Unexpected semimetallic {B}i{S}$_2$ at high pressure and high
  temperature.
\newblock {\em J. Phys. Chem. Lett.}, 9(19):5785, 2018.

\bibitem{hohenberg-pr-136-1964}
P.~Hohenberg and W.~Kohn.
\newblock {I}nhomogeneous electron gas.
\newblock {\em Phys. Rev.}, 136:B864, 1964.

\bibitem{kresse-cms-6-1996}
G.~Kresse and J.~Furthm\"{u}ller.
\newblock {E}fficiency of ab-initio total energy calculations for metals and
  semiconductors using a plane-wave basis set.
\newblock {\em Comput. Mat. Sci.}, 6:15, 1996.

\bibitem{perdew-prl-100-2008}
J.~P. Perdew, A.~Ruzsinszky, G.~I. Csonka, O.~A. Vydrov, G.~E. Scuseria, L.~A.
  Constantin, X.~Zhou, and K.~Burke.
\newblock {R}estoring the density-gradient expansion for exchange in solids and
  surfaces.
\newblock {\em Phys. Rev. Lett.}, 100:136406, 2008.

\bibitem{perdew-prl-102-2009}
J.~P. Perdew, A.~Ruzsinszky, G.~I. Csonka, O.~A. Vydrov, G.~E. Scuseria, L.~A.
  Constantin, X.~Zhou, and Kieron Burke.
\newblock {E}rratum: Restoring the density-gradient expansion for exchange in
  solids and surfaces [{P}hys. {R}ev. {L}ett. 100, 136406 (2008)].
\newblock {\em Phys. Rev. Lett.}, 102:039902, 2009.

\bibitem{monkhorst-prb-13-1976}
H.~J. Monkhorst and J.~D. Pack.
\newblock {S}pecial points for {B}rillouin-{Z}one integrations.
\newblock {\em Phys. Rev. B}, 13:5188, 1976.

\bibitem{PhysRevLett.106.145501.2011}
L.~Zhu, H.~Wang, Y.~Wang, J.~Lv, M.~Yanmei, Q.~Cui, Y.~Ma, and G.~Zou.
\newblock Substitutional alloy of {B}i and {T}e at high pressure.
\newblock {\em Phys. Rev. Lett.}, 106:145501, 04 2011.

\bibitem{JPhysChemC.117.10045.2013}
G.~Liu, L.~Zhu, M.~Yanmei, C.~Lin, J.~Liu, and Y.~Ma.
\newblock Stabilization of 9/10-fold structure in bismuth selenide at high
  pressures.
\newblock {\em J. Phys. Chem. C}, 117:10045, 05 2013.

\bibitem{murnaghan-pnas-30-1944}
F.~D. Murnaghan.
\newblock {T}he compressibility of media under extreme pressures.
\newblock {\em Proc. Natl. Acad. Sci.}, 30:244, 1944.

\bibitem{birch-pr-71-1947}
F.~Birch.
\newblock {F}inite elastic strain of cubic crystals.
\newblock {\em Phys. Rev.}, 71:809, 1947.

\bibitem{togo-prb-78-2008}
A.~Togo, F.~Oba, and I.~Tanaka.
\newblock {F}irst-principles calculations of the ferroelastic transition
  between rutile-type and {C}a{C}l$_2$-type {S}i{O}$_2$ at high pressures.
\newblock {\em Phys. Rev. B}, 78:134106, 2008.

\bibitem{chaput-prb-84-2001}
L.~Chaput, A.~Togo, I.~Tanaka, and G.~Hug.
\newblock {P}honon-phonon interactions in transition metals.
\newblock {\em Phys. Rev. B}, 84:094302, 2001.

\bibitem{PhysRevB.50.13035R.1994}
X.~Gonze, J.-C. Charlier, D.C. Allan, and M.P. Teter.
\newblock Interatomic force constants from first principles: The case of
  \ensuremath{\alpha}-quartz.
\newblock {\em Phys. Rev. B}, 50:13035, Nov 1994.

\bibitem{PhysRevB.55.10355.1997}
X.~Gonze and C.~Lee.
\newblock Dynamical matrices, born effective charges, dielectric permittivity
  tensors, and interatomic force constants from density-functional perturbation
  theory.
\newblock {\em Phys. Rev. B}, 55:10355, Apr 1997.

\bibitem{gajdos-prb-73-2006}
M.~Gajdo\v{s}, K.~Hummer, G.~Kresse, J.~Furthm\"{u}ller, and F.~Bechstedt.
\newblock {L}inear optical properties in the {PAW} methodology.
\newblock {\em Phys. Rev. B}, 73:045112, 2006.

\bibitem{PhysChemChemPhys.19.12452.2017}
J.~Skelton, L.~Burton, A.~Jackson, F.~Oba, S.~Parker, and A.~Walsh.
\newblock Lattice dynamics of the tin sulphides {S}n{S}$_2$, {S}n{S} and
  {S}n$_2${S}$_3$: Vibrational spectra and thermal transport.
\newblock {\em Phys. Chem. Chem. Phys.}, 19:12452, 05 2017.

\bibitem{PhonoptSpec}
J.~M. Skelton.
\newblock Phonopy-spectroscopy.

\bibitem{PhysRevB.91.094306.2015}
A.~Togo, L.~Chaput, and I.~Tanaka.
\newblock Distributions of phonon lifetimes in brillouin zones.
\newblock {\em Phys. Rev. B}, 91:094306, Mar 2015.

\bibitem{PhysRevB.65.104104.2002}
Y.~Page and P.~Saxe.
\newblock Symmetry-general least-squares extraction of elastic data for
  strained materials from ab initio calculations of stress.
\newblock {\em Phys. Rev. B}, 65:104104, 02 2002.

\bibitem{JPhysCondensMatter.28.275201.2016}
R.~Gaillac, P.~Pullumbi, and F.-X. Coudert.
\newblock {ELATE}: An open-source online application for analysis and
  visualization of elastic tensors.
\newblock {\em J. Phys. Condens. Matter}, 28:275201, 02 2016.

\bibitem{JSolidStateChem.213.116.2014}
J.~J. Carey, J.~P. Allen, D.~O. Scanlon, and G.~W. Watson.
\newblock The electronic structure of the antimony chalcogenide series:
  Prospects for optoelectronic applications.
\newblock {\em J. Solid State Chem.}, 213:116, 2014.

\bibitem{ChemSci.6.5255.2015}
V.~L. Deringer, R.~P. Stoffel, M.~Wuttig, and R.~Dronskowski.
\newblock Vibrational properties and bonding nature of {S}b$_2${S}e$_3$ and
  their implications for chalcogenide materials.
\newblock {\em Chem. Sci.}, 6:5255, 2015.

\bibitem{SolidStateSci.14.1211.2012}
H.~Koc, A.~M. Mamedov, E.~Deligoz, and H.~Ozisik.
\newblock First principles prediction of the elastic, electronic, and optical
  properties of {S}b$_2${S}$_3$ and {S}b$_2${S}e$_3$ compounds.
\newblock {\em Solid State Sci.}, 14(8):1211, 2012.

\bibitem{EJChem.6.S147.2009}
N.~Kuganathan.
\newblock Antimony selenide crystals encapsulated within single walled carbon
  nanotubes-a {DFT} study.
\newblock {\em E-J. Chem.}, 6:S147, 11 2009.

\bibitem{ZKristall.171.261.1985}
G.~P. {Voutsas}, A.~G. {Papazoglou}, P.~J. {Rentzeperis}, and D.~{Siapkas}.
\newblock {The crystal structure of antimony selenide, {S}b$_2${S}e$_3$}.
\newblock {\em Z. Kristall.}, 171(3-4):261, Jan 1985.

\bibitem{PhysicaB.406.287.2011}
T.~B.~Nasr, H.~Maghraoui-Meherzi, H.~Abdallah, and R.~Bennaceur.
\newblock Electronic structure and optical properties of {S}b$_2${S}$_3$
  crystal.
\newblock {\em Physica B}, 406:287, 01 2011.

\bibitem{PhysChemChemPhys.16.345.2014}
Y.~Liu, K.~Chua, T.~C. Sum, and C.~Gan.
\newblock First-principles study of the lattice dynamics of {S}b$_2${S}$_3$.
\newblock {\em Phys. Chem. Chem. Phys.}, 16:345, 11 2013.

\bibitem{PhysChemMinerals.30.463.2003}
L.~Lundegaard, R.~Miletich, T.~Balic-Zunic, and E.~Makovicky.
\newblock Equation of state and crystal structure of {S}b$_2${S}$_3$ between 0
  and 10 {GP}a.
\newblock {\em Phys. Chem. Minerals}, 30:463, 09 2003.

\bibitem{ZKristall.135.308.1972}
P.~Bayliss and W.~Nowacki.
\newblock Refinement of the crystal structure of stibnite, {S}b$_2${S}$_3$.
\newblock {\em Z. Kristall.}, 135:308, 2015.

\bibitem{PhysChemMiner.135.29.254.2002}
A.~Kyono, M.~Kimata, M.~Matsuhisa, Y.~Miyashita, and K.~Okamoto.
\newblock Low-temperature crystal structures of stibnite implying orbital
  overlap of sb 5$^2$ inert pair electrons.
\newblock {\em Phys. Chem. Miner.}, 29(4):254, May 2002.

\bibitem{JMolModel.20.2180.2014}
H.~Koc, H.~Ozisik, E.~Delig{\"o}z, A.~M. Mamedov, and E.~Ozbay.
\newblock Mechanical, electronic, and optical properties of {B}i$_2${S}$_3$ and
  {B}i$_2${S}e$_3$ compounds: first principle investigations.
\newblock {\em J. Mol. Model.}, 20(4):2180, Mar 2014.

\bibitem{CompMatSci.101.301.2015}
E.~Zahedi and B.~Xiao.
\newblock {DFT} study of structural, elastic properties and thermodynamic
  parameters of {B}i$_2${S}$_3$ under hydrostatic pressures.
\newblock {\em Comp. Mat. Sci.}, 101:301, 2015.

\bibitem{PhysChemMinerals.32.578.2005}
L.~Lundegaard, E.~Makovicky, T.~Ballaran, and T.~Balic-Zunic.
\newblock Crystal structure and cation lone electron pair activity of
  {B}i$_2${S}$_3$ between 0 and 10 gpa.
\newblock {\em Phys. Chem. Minerals}, 32:578, 12 2005.

\bibitem{JChemPhys.128.084714.2008}
A.~E. Mattsson, R.~Armiento, J.~Paier, G.~Kresse, J.~M. Wills, and T.~R.
  Mattsson.
\newblock The {AM05} density functional applied to solids.
\newblock {\em J. Chem. Phys.}, 128(8):084714, 2008.

\bibitem{PhysRevB.72.085108.2005}
R.~Armiento and A.~E. Mattsson.
\newblock Functional designed to include surface effects in self-consistent
  density functional theory.
\newblock {\em Phys. Rev. B}, 72:085108, Aug 2005.

\bibitem{PhysRevB.79.155101.2009}
A.~E. Mattsson and R.~Armiento.
\newblock Implementing and testing the {AM05} spin density functional.
\newblock {\em Phys. Rev. B}, 79:155101, Apr 2009.

\bibitem{HighTempHighPress.43.351.2014}
D.~Fan, J.~Xu, J.~Liu, Y.~Li, and H.~Xie.
\newblock Thermal equation of state of natural stibnite up to 25.7 {GP}a and
  533 {K}.
\newblock {\em High Temp. High Press.}, 43:351, 01 2014.

\bibitem{InorChem.50.11291.2011}
J.~Zhao, H.~Liu, L.~Ehm, Z.~Chen, S.~Sinogeikin, Y.~Zhao, and G.~Gu.
\newblock Pressure-induced disordered substitution alloy in {S}b$_2${T}e$_3$.
\newblock {\em Inor. Chem.}, 50(22):11291, 2011.

\bibitem{ProcCambridgePhilosSoc36.160.1940}
M.~Born.
\newblock On the stability of crystal lattices. {I}.
\newblock {\em Proc. Cambridge Philos. Soc.}, 36:160, 1940.

\bibitem{Dove.IntLattDyn}
Martin~T. Dove.
\newblock {\em Introduction to Lattice Dynamics}.
\newblock Cambridge University Press, 1993.

\bibitem{Dove.StrutDyn}
Martin~T. Dove.
\newblock {\em Structure and Dynamics: An Atomic View of Materials}.
\newblock Oxford Master Series in Physics, 2003.

\bibitem{AmMin.82.213.1997}
M.~T. Dove.
\newblock Review: Theory of displacive phase transitions in minerals.
\newblock {\em Am. Min.}, 82:213, 2015.

\bibitem{BullMaterSci.1.129.1979}
G.~Venkataraman.
\newblock Soft modes and structural phase transitions.
\newblock {\em Bull. Mater. Sci.}, 1(3):129, Dec 1979.

\bibitem{PhysRevLett.111.025503.2013}
M.~Di~Gennaro, S.~Saha, and M.~Verstraete.
\newblock Role of dynamical instability in the ab initio phase diagram of
  calcium.
\newblock {\em Phys. Rev. Lett.}, 111:025503, 07 2013.

\bibitem{Seeger.Wiley.2007}
H.-J. Quadbeck-Seeger.
\newblock {\em World of the Elements: Elements of the World}.
\newblock Wiley, 2007.

\bibitem{JChemPhys.123.204708.2005}
M.~Sternik and K.~Parlinski.
\newblock {F}ree-energy calculations for the cubic {Z}r{O}$_2$ crystal as an
  example of a system with a soft mode.
\newblock {\em J. Chem. Phys.}, 123:204708, 2005.

\bibitem{PhysRevB.91.144107.2015}
E.~Lora da~Silva, J.~M. Skelton, S.~C. Parker, and A.~Walsh.
\newblock Phase stability and transformations in the halide perovskite
  {C}s{S}n{I}$_{3}$.
\newblock {\em Phys. Rev. B}, 91:144107, Apr 2015.

\bibitem{JPhysCondensMatter.9.8579.1997}
B~B Karki, G~J Ackland, and J~Crain.
\newblock Elastic instabilities in crystals from ab initio stress - strain
  relations.
\newblock {\em J. Phys. Condens. Matter.}, 9(41):8579, oct 1997.

\bibitem{PhysRevB.90.224101.2014}
F.~Mouhat and F.-X. Coudert.
\newblock Necessary and sufficient elastic stability conditions in various
  crystal systems.
\newblock {\em Phys. Rev. B}, 90:224104, Dec 2014.

\bibitem{pfeiffer.PhDthesis.2009}
S~Pfeiffer.
\newblock {\em Darstellung neuer Alkalioxomanganate über die Azid/Nitrat-Route
  und Hochdruck-Hochtemperaturuntersuchungen an Sulfiden und Seleniden von
  Elementen der 14. und 15. Gruppe}.
\newblock PhD thesis, Universit\"{a}t Stuttgart, 2009.

\bibitem{PhysRevB.84.184110.2011}
R.~Vilaplana, D.~Santamar\'{\i}a-P\'erez, O.~Gomis, F.~J. Manj\'on,
  J.~Gonz\'alez, A.~Segura, A.~Mu\~noz, P.~Rodr\'{\i}guez-Hern\'andez,
  E.~P\'erez-Gonz\'alez, V.~Mar\'{\i}n-Borr\'as, V.~Mu\~noz Sanjose, C.~Drasar,
  and V.~Kucek.
\newblock Structural and vibrational study of {B}i$_2${S}e$_3$ under high
  pressure.
\newblock {\em Phys. Rev. B}, 84:184110, Nov 2011.

\bibitem{InorgChem.57.8241.2018}
R.~Vilaplana, S.~G. Parra, A.~Jorge-Montero, P.~Rodríguez-Hernández,
  A.~Munoz, D.~Errandonea, A.~Segura, and F.~J. Manj\'on.
\newblock Experimental and theoretical studies on $\alpha$-{I}n$_2${S}e$_3$ at
  high pressure.
\newblock {\em Inorg. Chem.}, 57(14):8241, 2018.

\end{thebibliography}
%\bibliographystyle{apsrev4-1}

\end{document}